\documentstyle[psfig,eqsecnum,prd,aps,preprint,tighten]{revtex}

 \newfont{\titlefont}{cmssbx10 scaled\magstep5}
\newcommand{\CA}{{\cal A}}
\newcommand{\CB}{{\cal B}}

\newcommand{\CD}{{\cal D}}

\newcommand{\CG}{{\cal G}}
\newcommand{\CH}{{\cal H}}

\newcommand{\CL}{{\cal L}}

\newcommand{\CN}{{\cal N}}
\newcommand{\CO}{{\cal O}}

\newcommand{\bea}{\begin{eqnarray}} \newcommand{\eea}{\end{eqnarray}}
\newcommand{\beq}{\begin{equation}} \newcommand{\eeq}{\end{equation}}
\newcommand{\non}{\nonumber} 
 \newcommand{\te}{temperature\ }
\newcommand{\lmk}{\left(} \newcommand{\rmk}{\right)}
\newcommand{\lkk}{\left[} \newcommand{\rkk}{\right]}
\newcommand{\lhk}{\left \{ } \newcommand{\rhk}{\right \} }
\newcommand{\del}{\partial} 
\newcommand{\vect}[1]{\mbox{\boldmath${#1}$}}
\newcommand{\bib}{\bibitem} \newcommand{\new}{\newblock}
\newcommand{\la}{\left\langle} \newcommand{\ra}{\right\rangle}
 
\newcommand{\gtilde} {~ \raisebox{-1ex}{$\stackrel{\textstyle >}{\sim}$} ~} 
\newcommand{\ltilde} {~ \raisebox{-1ex}{$\stackrel{\textstyle <}{\sim}$} ~}
 
\newcommand{\phip}{\phi_+}

\begin{document}

\begin{flushright}
  UTAP-247 \\ YITP-96-61
\end{flushright}

\begin{center}
  {\Large \bf Numerical approach to the onset of the electroweak
    phase transition \\} 
  \vskip 1cm {\large Masahide Yamaguchi} \\ 
  \vskip 0.2cm {\large \em Department of Physics, University of Tokyo} 
  \\ 
  \vskip 0.1cm {\large \em Tokyo 113, Japan} 
  \vskip 0.5cm {\large Jun'ichi Yokoyama} \\ 
  \vskip 0.2cm {\large \em Yukawa Institute
    for Theoretical Physics, Kyoto University} \\ 
  \vskip 0.1cm {\large \em Kyoto 606-01, Japan}
\end{center}


\begin{abstract}

  We investigate whether the universe was homogeneously in the false
  vacuum state at the critical temperature of a weakly first-order
  phase transition such as the electroweak phase transition in terms
  of a series of numerical simulations of a phenomenological Langevin
  equation, whose noise term is derived from the effective action but
  the dissipative term is set so that the fluctuation-dissipation
  relation is met. The correlation function of the noise terms given
  by a non-equilibrium field theory has a distinct feature if it
  originates from interactions with a boson or with a fermion. The
  spatial correlation function of noises from a massless boson damps
  with a power-law, while the fermionic noises always damp
  exponentially above the inverse-temperature scale. In the simulation
  with one-loop effective potential of the Higgs field, the latter
  turns out to be more effective to disturb the homogeneous field
  configuration. Since noises of the both types are present in the
  electroweak phase transition, our results suggest that conventional
  picture of a phase transition, namely, nucleation of critical
  bubbles in a homogeneous background does not apply or the one-loop
  approximation breaks down in the standard model.

\end{abstract}

\thispagestyle{empty} \setcounter{page}{0} \newpage
\setcounter{page}{1}

\section{Introduction}

\label{sec:introduction}

\indent

In modern cosmology our universe has presumably experienced a number
of phase transitions in the early stage of its evolution.  Among them
those at the grand unification scale, which may be related with
inflation and/or formation of topological defects, are very
speculative in that we know neither the symmetry-breaking pattern nor
the initial state before the transition which must be highly
non-thermal due to the rapid cosmic expansion then. In contrast, the
dynamics of the electroweak phase transition (EWPT) in the standard
model is a much more solid subject of study because we have a
perturbative expression of the effective potential with only one
undetermined degree of freedom, namely, the Higgs mass, $M_{H}$,
\cite{AH} \cite{She} \cite{DLHLL} and we may well assume the thermal
state before the transition since the relevant particle interaction
rates are much larger than the cosmic expansion rate by that time.

Nevertheless the dynamics of EWPT is not fully understood yet, mainly
because, although one-loop effective potential of the Higgs field,
$\phi$, shows it is of first order, the potential barrier between the
two minima at the critical temperature, $T_c$, is so shallow that it
has been doubted if the conventional picture of nucleation of critical
bubbles in the homogeneous false-vacuum background really works.
Whether the transition is of first order with super-cooling is a very
important cosmological issue to judge if electroweak baryogenesis
\cite{CKN} is possible.

Much work has already been done on this topic. For example, 
Gleiser, Kolb, and Watkins \cite{GKW}
considered the role of subcritical bubbles of the correlation volume
as a noise effect in a weakly first-order phase transition, and
Gleiser and Kolb \cite{GK} concluded that for the Higgs mass larger
than 57GeV, the universe is not in the false vacuum state uniformly
with $\phi=0$ but in a mixture of $\phi=0$ and the true vacuum
$\phi\equiv\phip(T_c)$\footnote{At the critical temperature, the
  notion of the false and the true vacua are not well defined since
  the both minima are degenerate.  Here we use the same terminology as
  in lower temperatures.}.  See also Gleiser and Ramos \cite{GRPL}.
Gleiser \cite{Gle} and Borrill and Gleiser \cite{BG} have confirmed
occurrence of such ``phase mixing'' by numerical simulation of a phase
transition, solving a simple phenomenological Langevin equation of the
form

\beq 
\Box \phi(\vect x,t)+\eta\dot{\phi}(\vect x,t)+ V'_{\rm
  eff}(\phi, T_c) = \xi(\vect x,t)\:, \label{eqn:simplelangevin} 
\eeq

\noindent
at the critical temperature $T = T_{c}$.  Here an overdot denotes time
derivation, $V_{\rm eff}(\phi,T_{c})$ is an effective potential, and $\xi
(\vect x,t)$ is a random Gaussian noise with the correlation function

\beq 
  \langle\,\xi (\vect x_1,t_1)\xi(\vect x_2,t_2)\,\rangle =
      \CD\delta(\vect x_1-\vect x_2) \delta(t_1-t_2)\:,
\label{eqn:whitenoise} 
\eeq

\noindent
satisfying the fluctuation-dissipation relation $\CD=2\eta T_c$.
Later Shiromizu {\it et al}.\ \cite{SMY} treated the size of a
subcritical bubble as a statistical variable and discussed its typical
size is smaller than the correlation length. They concluded that for
any experimentally allowed value of the Higgs mass, or $M_{H} \gtilde
60$GeV \cite{dat}, the phase mixing does occur already at the critical
temperature.  Furthermore, in the
Monte-Carlo lattice simulations  of the Euclidean 
four dimensional theory or the reduced three dimensional model of the 
finite-temperature electroweak theory, analytical
cross-over behavior is observed for most values of the Higgs mass permitted
from the experiment \cite{Ja} \cite{Ru}. 

On the other hand, Dine {\it et al}. \cite{DLHLL} calculated
root-mean-square amplitude of the Higgs field on the correlation
scale, or the inverse-mass length scale at $\phi = 0$ at the critical
temperature and concluded that for the Higgs boson with $M_{H} \simeq
60$GeV it is much smaller than the distance between the two minima and
the fraction of the asymmetric phase of the Universe is negligible
($e^{-12}$) so that subcritical fluctuations do not affect the
dynamics of EWPT. Bettencourt \cite{Bet} confirmed their conclusion by
estimating the probability that the mean value of $\phi$ averaged over
a correlation volume is larger than the distance to the maximum of the
effective potential separating the two minima and finding that it is
extremely small.  Finally in response to Shiromizu {\it et al}.\cite{SMY},
Enqvist {\it et al}.\cite{ERV} treated both the amplitude and the spatial
size of subcritical fluctuations as statistical variables and
discussed that subcritical bubbles, if they exist at all, resemble the
critical bubbles and that the usual description of a first-order phase
transition applies.  Their analysis, however, has a problem that it
suffers from severe divergence and they had to introduce cut off {\it
ad hoc}.

Thus a number of independent analyses have drawn different conclusions
about how EWPT proceeds.  In the present paper we attempt to elucidate
why such discrepancy has arisen with the help of numerical simulations
of a phenomenological Langevin equation which is better motivated than
(\ref{eqn:simplelangevin}) and (\ref{eqn:whitenoise}) from a
non-equilibrium field theory. In fact the origin of the discrepancy is
quite simple: it only reflects at which spatial scale one estimates
the amplitude of fluctuations. We try to approach the problem step by
step.

The rest of the paper is organized as follows.  We start with a
re-analysis of the simple Langevin equation (\ref{eqn:simplelangevin})
with white noise (\ref{eqn:whitenoise}) in Sec.\ \ref{sec:re-analysis}.
First we consider a non-selfinteracting massive scalar model and show
that numerical calculations of (\ref{eqn:simplelangevin}) and
(\ref{eqn:whitenoise}) on a lattice can reproduce the
finite-temperature spatial correlation function of a massive scalar
field correctly as long as we take the lattice spacing comfortably
smaller than the correlation length or the Compton wave length.  We
then solve the same equation but with an effective potential in the
standard model as was done by Borrill and Gleiser \cite{BG}. We find
not only the behavior of the correlation function but also the limit
on $M_{H}$ above which phase mixing occurs change drastically
depending on the lattice spacing.  In order to obtain a sensible bound
on $M_{H}$, therefore, we should choose a reasonable value of the
lattice spacing with the help of a fundamental theory.  This issue is
discussed in Sec.\ \ref{sec:properties} and a new phenomenological
Langevin equation is proposed.  In Sec.\ \ref{sec:numerical} we report
the results of numerical simulations of the dynamics of the field based
on this equation. In Sec.\ \ref{sec:analytic} we give an intuitive
explanation for the numerical results by using a simple Boltzmann
equation. Finally Sec.\ \ref{sec:summary} is devoted to summary and
discussion.

\section{Re-analysis of the simple Langevin equation}

\label{sec:re-analysis}

\indent

In this section we elucidate the origin of the discrepancy in the
previous literatures. For this purpose, we first solve the simple
Langevin equation (\ref{eqn:simplelangevin}) in the case only the mass
term is present in the potential, namely,
\beq
   (\Box+m^2)\,\phi(\vect x,t)+\eta\dot{\phi}(\vect x,t)
      = \xi(\vect x,t)\:, \label{eqn:simplelangevin2} 
\eeq 
where $\xi (\vect x,t)$ is a random Gaussian noise satisfying
(\ref{eqn:whitenoise}) with $\CD=2\eta T$.

After discretizing the system on a lattice, we follow the time
evolution from the initial condition, $\phi(\vect x,0) = 0$ and
$\dot\phi(\vect x,0) = 0$, on each lattice point. The dimensionless
variables $\tilde{\vect x} \equiv T \vect x,\,\tilde t \equiv
Tt,\,\Phi \equiv T^{-1}\phi,\, \tilde\eta \equiv T^{-1}\eta,\,
\tilde\xi \equiv T^{-3}\xi$, and $\mu \equiv T^{-1}m$ are introduced
for numerical calculations, but we omit the tildes below. We arrange
three different lattices for comparison. One has the total lattice
points, $N = 32^3$, grid spacing, $\delta x = 1.0$, time step, $\delta
t = 0.1$, and total run time, $t = 500$, another has $N =
40^3,\,\delta x = 0.8,\,\delta t = 0.1$, and $t = 500$, and the other
has $N = 64^3,\,\delta x = 0.5,\,\delta t = 0.1$, and $t = 500$. Using
the second order staggered leapfrog method, the discretized master
equation reads
\bea 
   \dot\Phi_{\vect i,n+1/2} &=& \frac{1}{1+\frac12\eta\,\delta t}
       \lkk\,
          \lmk\,1-\frac12\eta\,\delta t\,\rmk \dot\Phi_{\vect i,n-1/2}
               +\delta t \lmk\,
                  \nabla^2\Phi_{\vect i,n}-\mu^2\Phi_{\vect i,n}
                     +\xi_{\vect i,n}
                         \,\rmk  
       \,\rkk \:, \non \\
   \Phi_{\vect i,n+1} &=& \Phi_{\vect i,n} 
         + \delta t\,\dot\Phi_{\vect i,n+1/2}
       \:, \non \\
   \nabla^2\Phi_{\vect i,n} &\equiv& \sum_{s = x, y, z} 
       \frac{\Phi_{i_{s}+1_{s},n} - 2\,\Phi_{i_{s},n} +
             \Phi_{i_{s}-1_{s},n}}{(\delta x)^2} \:,
\eea   
where $\vect i$ represents spatial index and $n$ temporal one, and
$\mu$ is set to be $0.125$. The correlation of the noise is given on
the lattice by
\beq
  \la\,\xi_{\vect i_{1},n_{1}}\xi_{\vect i_{2},n_{2}}\,\ra
      =2\eta\frac{1}{\delta t}\delta_{n_{1},n_{2}}
            \frac{1}{(\delta x)^3}\delta_{\vect i_{1},\vect i_{2}} \:.
\eeq
Since it is white both spatially and temporally, we have only to
generate Gaussian white noise on each grid,
\beq
   \xi_{\vect i,n} = \sqrt{\frac{2\eta}{\delta t (\delta x)^3}}
                       \,\CG_{\vect i,n}
        \:,        \label{eqn:Gaussiannoise}
\eeq
where $\CG_{\vect i,n}$ is a Gaussian random number with a vanishing
mean and a unit dispersion. The periodic boundary condition is
imposed. Under the above conditions we take the ensemble average over
five different noises for each case.  The correlation function $C(r),
\equiv \la\,\Phi(\vect x)\Phi(\vect y)\,\ra$ with $r = |\vect x-\vect
y|$, is numerically obtained by calculating all the combinations of
the product $\Phi(\vect x)\Phi(\vect y)$ satisfying $r-0.5 \le |\vect
x-\vect y|<r+0.5$.

On the other hand, the analytic expression of the equal-time
correlation function of a non-selfinteracting scalar field with mass
$m$ at temperature $T$ is given by

\bea 
   \la\,\phi(\vect x)\phi(\vect y)\,\ra &=&
         \frac{1}{\beta}\sum_{n=-\infty}^{\infty}\int
           \frac{d^3\vect k}{(2\pi)^3} \frac{1}{\omega_{n}^2+k^2+m^2} \, 
             e^{i \vect{k}\,\cdot\,(\vect{x-y})}  
                 \non   \\ &=&
          \frac{m}{4\pi^{2}r} 
               \lkk\,K_{1}(mr)+2\sum_{n=1}^{\infty}
                   \frac{r}{\sqrt{r^2+n^2 \beta^2}} 
                     \,K_{1} \lmk\, m\sqrt{r^2+n^2 \beta^2}\,\rmk
                \,\rkk\:,  \label{eqn:nonself}
\eea

\noindent
where $r=|\vect x-\vect y|$, $\beta=1/T$, $\omega_{n}=2\pi n/\beta$,
and $K_{j}$ is the modified Bessel function of the $j$-th order. This
correlation function damps exponentially above the inverse-mass scale.

\begin{figure}[htb]
\begin{center}
\leavevmode\psfig{figure=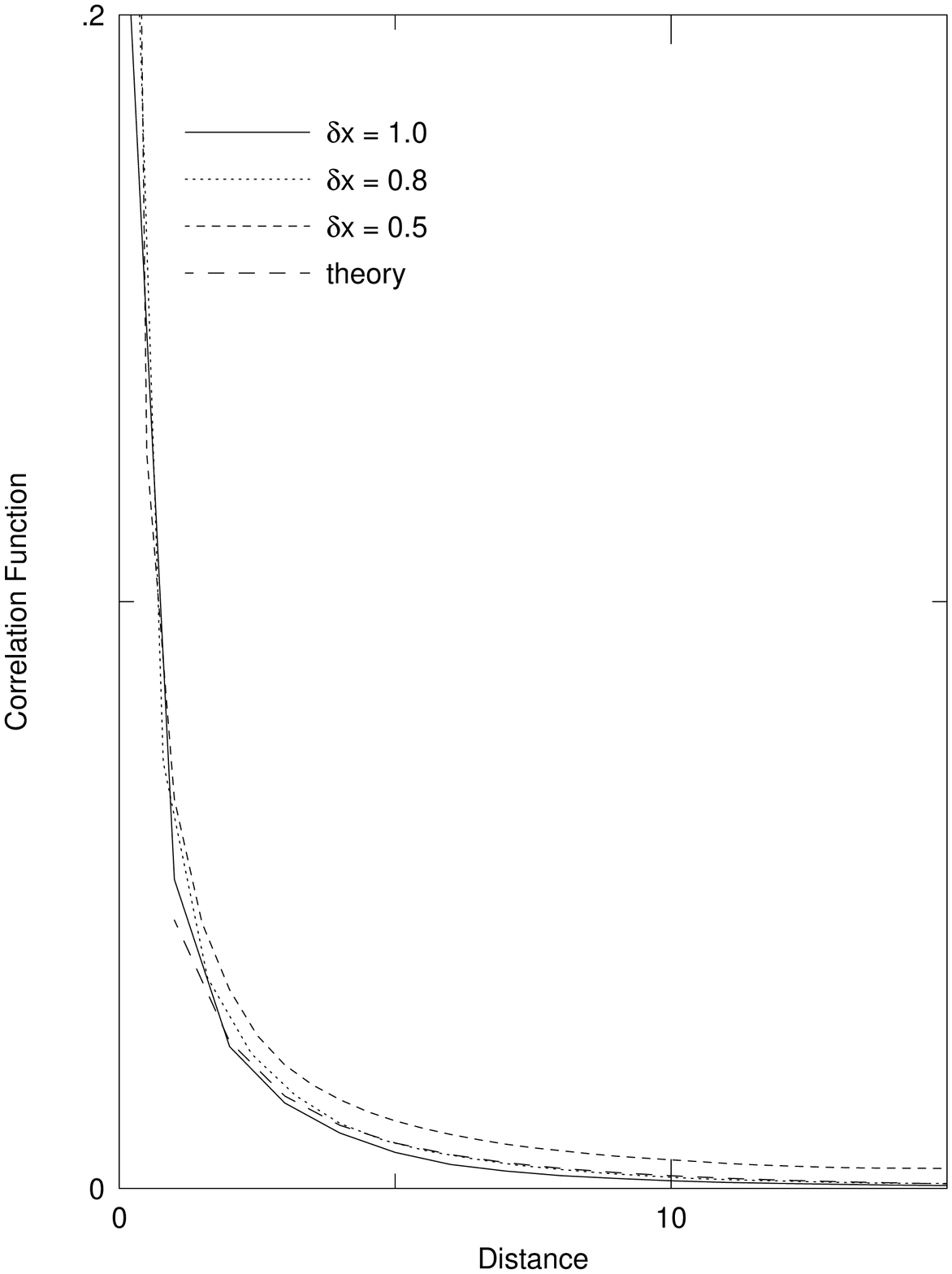,height=20cm}
\end{center}
\caption{Correlation function $\la \Phi(\vect x)\Phi(\vect y) \ra$ for
the massive non-selfinteracting field at finite temperature.}
\label{fig:nonself}
\end{figure}

Both numerical and analytic results are depicted in Fig.\
\ref{fig:nonself} in the dimensionless unit \footnote{The correlation
function derived analytically is reguralized by subtraction of the
vacuum energy.}. We find that the correlation functions obtained from
numerical calculation damp in a manner independent of the lattice
spacings and also that they coincide with the analytic formula
(\ref{eqn:nonself}), namely, exponentially damp with the correlation
length of the inverse-mass. Thus for the case of the massive
non-selfinteracting scalar field, not only the simple Langevin
equation (\ref{eqn:simplelangevin}) with the random noise
(\ref{eqn:whitenoise}) but also its numerical solution on a lattice
can reproduce its actual finite-\te behavior.

Next we consider an interacting scalar field borrowing the one-loop
improved effective potential, $V_{EW}$, of the Higgs field in the
electroweak theory following Borrill and Gleiser \cite{BG}. The master
equation is given by
\beq 
   \Box \phi(\vect x,t)+\eta\dot{\phi}(\vect x,t)+ V'_{EW}(\phi, T_c) 
      = \xi(\vect x,t) \:, \label{eqn:simplelangevin3} 
\eeq 
where $\xi (\vect x,t)$ is again a random Gaussian noise with no
correlation.
$V_{EW}$ is given in \cite{AH}\cite{She}\cite{DLHLL} as
\beq
  V_{EW}(\phi,T)
       = D(T^2-T_{2}^2)\phi^2-ET\phi^3+\frac14\lambda_{T}\phi^4 
           \:,   \label{eqn:potential}
\eeq
where 
\bea 
  D &=&
     \frac{1}{24}
      \lkk\,6\lmk\frac{M_{W}}{\sigma}\rmk^2
        + 3\lmk\frac{M_{Z}}{\sigma}\rmk^2
             + 6\lmk\frac{M_{t}}{\sigma}\rmk^2\,\rkk 
          = 0.169  \:, \\ 
  E &=& 
     \frac{1}{12\pi}
      \lkk\,6\lmk\frac{M_{W}}{\sigma}\rmk^3
             +3\lmk\frac{M_{Z}}{\sigma}\rmk^3\,\rkk 
          = 0.00965  \:,
\eea 
for $M_{W} = 80.6$ GeV, $M_{Z} = 91.2$ GeV, $M_{t} = 174$ GeV, and
$\sigma = 246$ GeV \cite{dat}. 
We also find
\bea
 T_{2} &=& \lkk \frac{M_{H}^2-8B\sigma^2}{4D}\rkk^{\frac{1}{2}}, \\
 M_{H}^2 &=& (2\lambda_{0} + 12B)\sigma^2, \\
 B &=& \frac{1}{64\pi^2\sigma^4}(6M_{W}^4+3M_{Z}^4-12M_{t}^4) = -0.00456
\eea
and the temperature-corrected Higgs self coupling is given by
\beq
  \lambda_{T} = \lambda_{0}
     -\frac{1}{16\pi^2} 
      \lkk\,\sum_{B}g_{B}\lmk \frac{M_{B}}{\sigma}\rmk^4
             \ln \lmk\frac{M_{B}^2}{c_{B}T^2}\rmk
           -\sum_{F}g_{F}\lmk \frac{M_{F}}{\sigma}\rmk^4
             \ln \lmk\frac{M_{F}^2}{c_{F}T^2}\rmk 
      \,\rkk \:,
\eeq
where the sum is performed over bosons and fermions with their degrees
of freedom $g_{B(F)}$ and $\ln c_{B}=5.41, \ln c_{F}=2.64$\,. The
Higgs field, $\phi$, appearing in the potential (\ref{eqn:potential})
in the actual electroweak theory is of course the amplitude of an
$SU(2)$ doublet complex scalar field. But in solving the Langevin
equation (\ref{eqn:simplelangevin3}), we neglect the gauge-nonsinglet
nature of the field for simplicity 
and constrain its dynamics along the real-neutral
component to treat $\phi$ as if it was a real singlet field as in
Borrill and Gleiser \cite{BG}. Thus this should not be regarded as
the simulation of the actual Higgs field, although our results would
be suggestive to it.

Introducing dimensionless variables, $\tilde{\vect x} \equiv
(2D)^{1/2}T_{2}
\vect x$, $\tilde t \equiv (2D)^{1/2}T_{2}t$, $\Phi \equiv
(2D)^{-1/4}T_{2}^{-1}\phi$, $\tilde\eta \equiv
(2D)^{-1/2}T_2^{-1}\eta$, $\tilde\xi \equiv (2D)^{-5/4}T_{2}^{-3}\xi$,
$\theta \equiv T/T_{2}$, and $V_{EW}(\phi) \equiv
(2D)^{3/2}T_{2}^4U(\Phi)$, Eqs.\,(\ref{eqn:simplelangevin3}) and
(\ref{eqn:potential}) reduce to

\beq 
  \tilde\Box \Phi(\tilde x) + \frac{\del U(\Phi)}{\del \Phi}
   +\eta\dot \Phi(\tilde{\vect x'},\tilde t) = \tilde\xi(\tilde x) 
   \;,
       \label{eqn:lesseq}
\eeq   
and
\beq
  U(\Phi)=\frac12(\theta^2-1)\Phi^2-\frac{\alpha}{3}\theta \Phi^3
        +\frac{\tilde\lambda}{4}\Phi^4 \:.
       \label{eqn:lesspote}
\eeq

\begin{figure}[htb]
\begin{center}
\leavevmode\psfig{figure=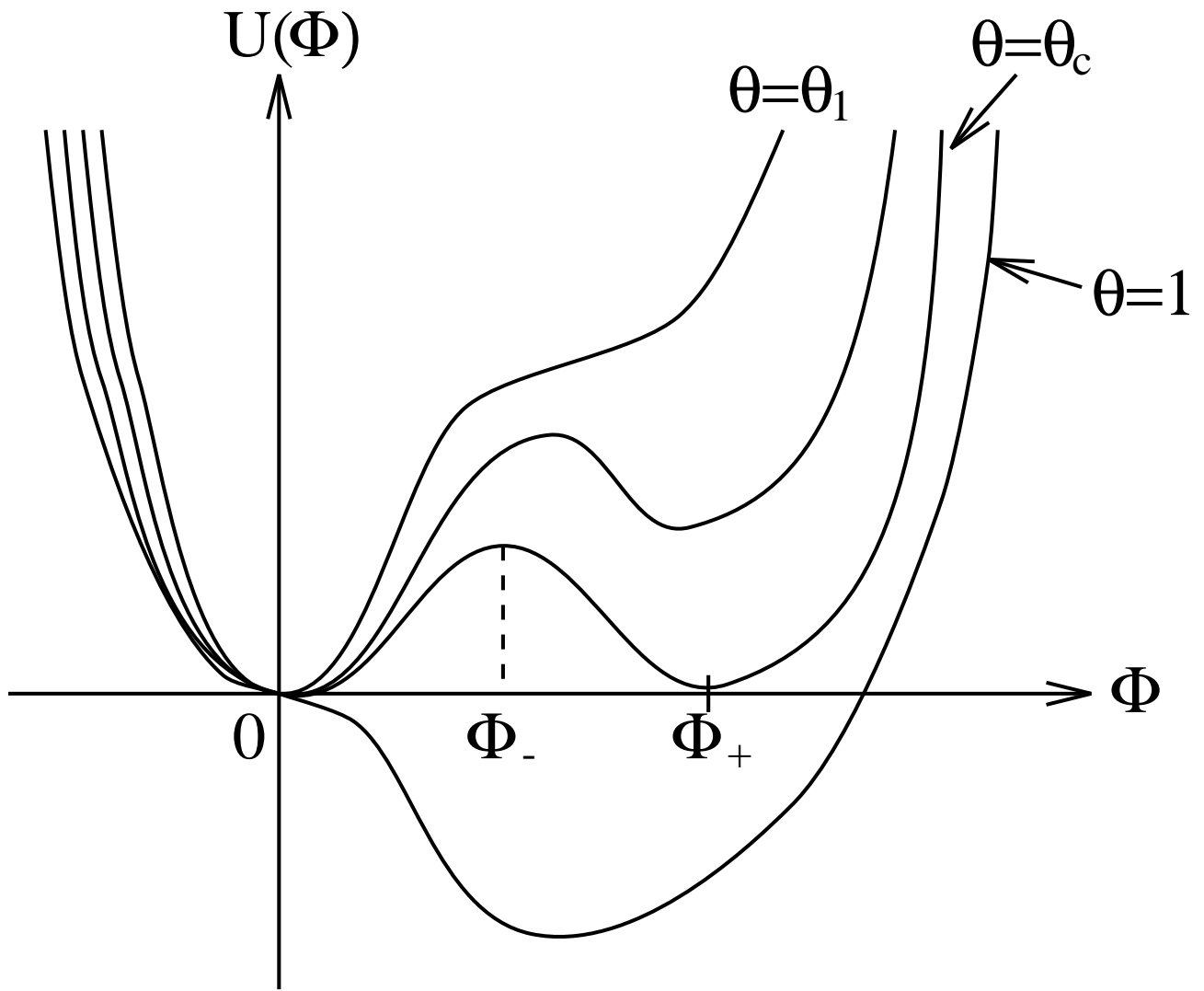,width=17cm}
\end{center}
\caption{One-loop improved effective potential of the Higgs field in
dimensionless unit.}
\label{fig:fig14}
\end{figure}

\noindent
Here dimensionless parameters are defined as
$\alpha=(2D)^{-3/4}(3E)=0.065,\,\tilde\lambda=(2D)^{-1/2}
\lambda_{T}=1.72\lambda_{T}$. 
Hereafter we omit the tildes. The effective potential is depicted in
Fig.\ \ref{fig:fig14}. For $\theta > \theta_{1} \equiv
(1-\alpha^2/4\lambda)^{-1/2}$ there is only one minimum at
$\Phi=0$. At $\theta=\theta_{1}$ appears the inflection point
$\Phi=\alpha\theta_{1}/2\lambda$. As the \te further drops, another
minimum, which is metastable, appears and at the critical temperature,
$\theta_{c} \equiv (1-2\alpha^2/9\lambda)^{-1/2}$, the two minima,
$\Phi=0$, and $ \Phi_{+} \equiv \frac{\alpha \theta}{2\lambda}
\lkk\,1+\sqrt{1-4\lambda(1-1/\theta^2)/\alpha^2}\,\rkk$, are
degenerate. Below $\theta=\theta_{c}$, the symmetric state $\Phi=0$
becomes metastable in turn and at $\theta_{2} \equiv 1$ disappears the
local maximum at $\Phi_{-} \equiv \frac{\alpha \theta}{2\lambda}
\lkk\,1-\sqrt{1-4\lambda(1-1/\theta^2)/\alpha^2}\,\rkk$.  This is a typical
model which represents a first-order phase transition.

We investigate whether the universe is in a homogeneous state of the
false vacuum at the onset of the phase transition or the critical
temperature following Borrill and Gleiser
\cite{BG}. Taking the initial condition as 
\beq
  \Phi(0, {\vect x}) = 0,~~~~~
   \dot \Phi(0, {\vect x}) = 0  \:,
\eeq

\noindent
for all ${\vect x}$\,, we follow the evolution of the field to trace
the fraction of the symmetric phase, $f_{0}(t)$, which is defined by
the fractional volume of the lattice space with $\Phi \leq \Phi_{-}$,
while the fraction of the asymmetric phase, $f_{+}(t) \equiv
1-f_{0}(t)$, is that with $\Phi \geq \Phi_{-}$. As in the
non-selfinteracting case, the system is discretized and the
second-order staggered leapfrog method is used.  The noises are also
generated in the same way. We have confirmed in all the cases of our
interest the change of $\eta$ affects only the relaxation time scale
but properties of the final configuration are insensitive to it
\cite{BG}.  So hereafter we report the results with $\eta=1$. Contrary
to Borrill and Gleiser \cite{BG}, we perform numerical calculations
with various values of the grid spacing, which turn out to affect the
final configuration greatly as seen below.

\begin{figure}[htb]
\begin{center}
\leavevmode\psfig{figure=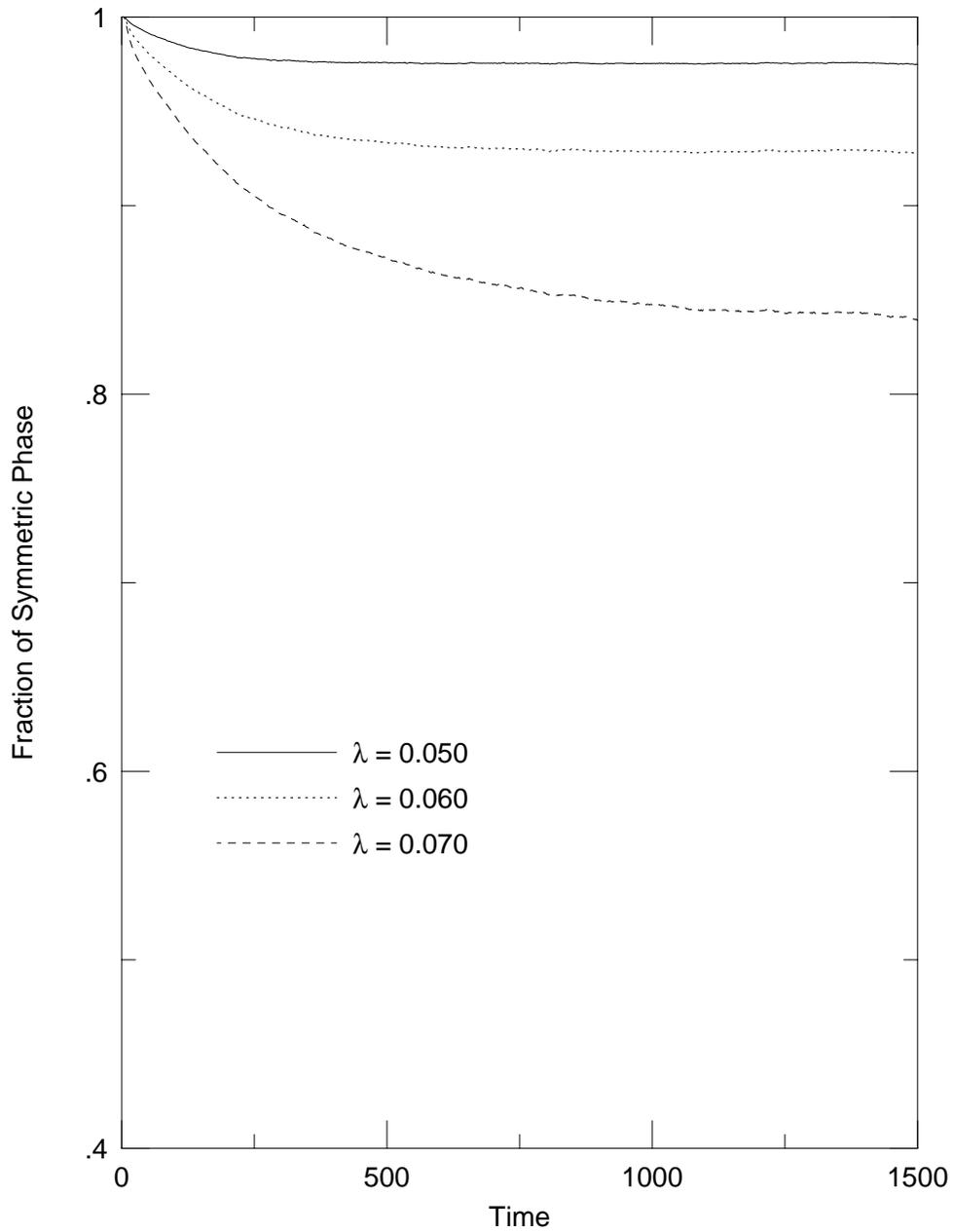,height=20cm}
\end{center}
\caption{Fraction of the symmetric phase, $f_{0}$, with the lattice
spacing $\delta x$ taken as the curvature scale of the potential at
$\Phi = 0$. ($\delta x = 7.2(\lambda = 0.050), 7.9(\lambda =
0.060), 8.6(\lambda = 0.070))$}  
\label{fig:curvature}
\end{figure}

\begin{figure}[htb]
\begin{center}
\leavevmode\psfig{figure=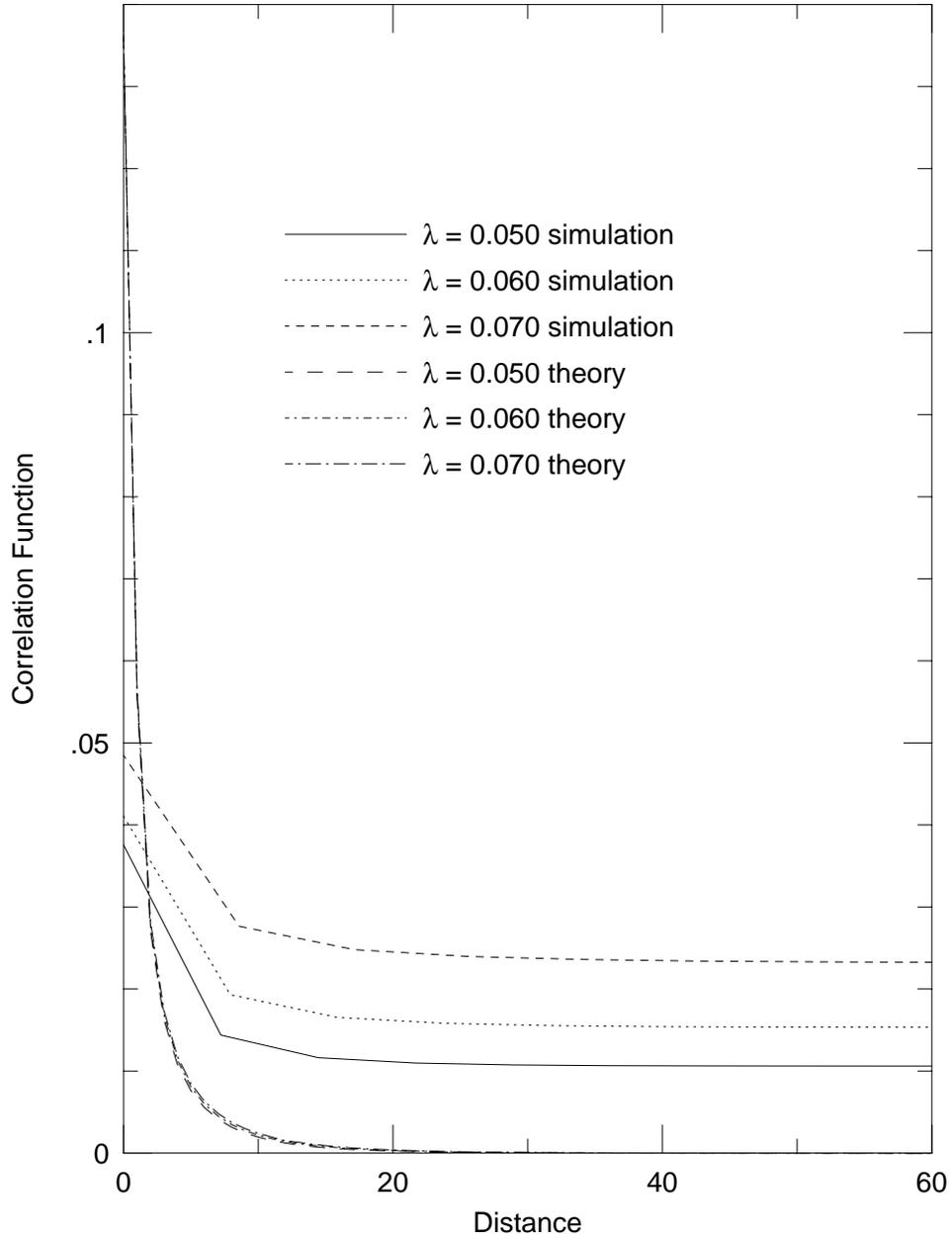,height=20cm}
\end{center}
\caption{Correlation function $\la \Phi(\vect x)\Phi(\vect y) \ra$ at
$t = 1500$ for the interacting scalar field with the lattice
spacing $\delta x$ taken as the curvature scale of the potential at
$\Phi = 0$.}  
\label{fig:ccurvature}
\end{figure}

First, as often assumed in the literatures
\cite{DLHLL} \cite{Bet}, we adopt so-called ``correlation length'' or
the curvature scale of the potential at $\Phi = 0$, $r \equiv
(U''[\Phi])^{-1/2}$, as the coarse-graining scale, namely, the lattice
spacing. We thus set $\delta x = (\theta_{c}^2-1)^{-1/2}$ and arrange
a lattice with $N = 64^3$, $\delta t = 0.1$, and $t = 1500\:.$ For
several $\lambda$'s, the fraction of the symmetric phase, $f_{0}(t)$
is depicted in Fig.\ \ref{fig:curvature}. For $\lambda = 0.06$
($\delta x \simeq 7.9$) corresponding to the Higgs mass $M_{H} \simeq
60$GeV \cite{dat}, the phase mixing does not occur. The result shows
that if the Higgs mass is not too large, the phase mixing does not
occur, which is consistent with the results of
\cite{DLHLL} \cite{Bet}. In Fig.\ \ref{fig:ccurvature} we have
depicted the correlation function, $\la \Phi(\vect x)\Phi(\vect y)
\ra$, at $t = 1500$, which does not necessarily approach zero for
large separation due to the fact that the average value of $\Phi$ is
not equal to zero.  In this figure we have also depicted theoretical
curves Eq.\ (\ref{eqn:nonself}) with $m = (U''[\Phi=0])^{1/2}$ in a 
dimensionless unit, namely,

\beq
   \la\,\Phi(\vect x)\Phi(\vect y)\,\ra =
                 \frac{m \sqrt{2D}}{2\pi^{2}} 
               \,\sum_{n=1}^{\infty}
                   \frac{1}{\sqrt{r^2+2Dn^2 / \theta_{c}^2}} 
                     \,K_{1} \lmk\, m \sqrt{
                       (r^2+2Dn^2 /\theta_{c}^2)}\,\rmk
                \,\:.  \label{eqn:correlation_cur}
\eeq

\noindent
The former damps much more mildly than the analytic counterpart
(\ref{eqn:correlation_cur}), motivating us to study the case with the 
smaller lattice spacing as has been done in \cite{BG}.

Next we investigate the dependence of results on the grid spacing
$\delta x$. Two lattices are arranged, one with $N = 64^3$, $\delta t
= 1.0$, $t = 1500$ and, $\delta x = 1.0$, and the other with the same
properties except for $\delta x = 0.5$.  The results are depicted in
Figs.\ 5(a) and 5(b). The former case
reproduces Borrill and Gleiser's result. Comparing both results, we
find that the smaller the lattice spacing becomes, the phase mixing
occurs for the smaller values of $\lambda$ corresponding to the
lighter Higgs mass. This result can be understood as follows. Taking
lattices is equivalent to cutting off the momentum. As is seen in
Eq.\ (\ref{eqn:Gaussiannoise}), the smaller lattice spacing we take,
the larger momentum dominates and the more easily the phase mixing
occurs. The correlation function are depicted in Figs.\ 6(a) and 6(b).
In the former case with $\delta x = 1.0$, it has a similar curve to
the analytic one (\ref{eqn:correlation_cur}) except for the offset,
while in the latter with $\delta x = 0.5$ the correlation damps more
rapidly. 

In order to examine dependence of the lattice spacing further, we have
also performed numerical simulations with $N = 64^3, \delta t = 0.1, t
= 3000$, and $\lambda$ fixed at $0.06$. As is seen in Fig.\
\ref{fig:lambda}, the results depend on the lattice spacings very
much. $\delta x \simeq 6.0$ is critical and if we take $\delta x$
smaller, the phase mixing is manifest. Therefore unless we specify the
lattice spacing from a physical argument, we cannot draw any
quantitative conclusion about whether two phases are mixed or
not. From the analogy with the non-selfinteracting massive scalar
field analyzed in the beginning of this section, many people have
adopted the Compton wavelength or the inverse-mass scale at $\Phi = 0$
as the coarse-graining scale. In the present case, however, the
correlation function changes significantly depending on the lattice
spacing as shown in Figs.\ 6(a) and 6(b) contrary to the case of
non-selfinteracting field. Thus the above results suggest that
the previous analyses adopting the
correlation length, or the inverse-mass at $\Phi = 0$, as the coarse
graining scale are inappropriate  in
the case of interacting potential.  Thus we must reconsider the
derivation of the Langevin equation before performing further
numerical analysis based on the simplified equation
(\ref{eqn:simplelangevin3}).

\section{Properties of the noises derived from a non-equilibrium
  quantum field theory}

\label{sec:properties}

\indent

Here we review a field theoretic approach to derive an effective
Langevin equation with particular emphasis on the origin of its noise
term.  The standard quantum field theory, which is appropriate for
evaluating the transition amplitude from an `in' state to an `out'
state for some field operator $\vect\CO$, $\la out|\,\vect\CO\,|in
\ra$, is not suitable to trace time evolution of an expectation value
in a non-equilibrium system. In order to follow the time development
of the expectation value of some fields, it is necessary to establish
an appropriate extension of the quantum field theory, which is often
called the in-in formalism. This was first done by Schwinger
\cite{Sch} and developed by Keldysh \cite{Kel}. Here, following
Morikawa \cite{Mor} and Gleiser and Ramos \cite{GR}, we first review
briefly the derivation of an effective Langevin-like equation for a
coarse-grained field using the non-equilibrium quantum field theory
based on the in-in formalism and then extract necessary information on
the noises which are essential for generating inhomogeneity of the
system.

\subsection{Non-equilibrium quantum field theory}

\indent

Let us consider the following Lagrangian density of a singlet scalar
field $\varphi$ interacting with another scalar field $\chi$ and a
fermion $\psi$ for illustration.
\beq 
   \CL = \frac12\,(\del_{\mu}\varphi)^2-\frac12\,m_{\varphi}^2\varphi^2 -
     \frac{1}{4!}\lambda\,\varphi^4
       + \frac12\,(\del_{\mu}\chi)^2-\frac12\,m_{\chi}^2\chi^2 -
     \frac{1}{4}\,g^2 \chi^2 \varphi^2 
       + i\bar\psi\gamma^{\mu}\del_{\mu}\psi - m_{\psi} \bar\psi \psi
     - f\varphi\bar \psi\psi \:.
\eeq 
Although the above Lagrangian density is much simpler than the
standard model, it turns out that this model fully accounts the nature
of bosonic noises arising from interactions with gauge particles and
Higgs self-interactions as well as fermionic noises from quarks and
leptons.
 
In order to follow the time development of $\varphi$, only the initial
condition is fixed, and so the time contour in a generating functional
starting from the infinite past must run to the infinite future
without fixing the final condition and come back to the infinite past
again. The generating functional is thus given by
\bea 
  Z[J,\eta,\bar\eta] &\equiv&
      \mbox{Tr} \lkk T\lmk \exp 
        \lkk i\int_{c} (J \varphi+K \chi +\eta\psi+\bar\eta \bar\psi 
          \rkk \rmk \rho\,\rkk \non \\ 
       &=& 
          \mbox{Tr} \lkk T_{-}\lmk
            \exp\lkk -i\int (J_{-}\varphi_{-}+K_{-}\chi_{-}+\eta_{-}\psi_{-}
                +\bar\eta_{-}\bar\psi_{-}) 
                    \rkk \rmk \right. \non \\
       && \qquad \qquad \qquad \left.   \times\, 
          T_{+}\lmk \exp\lkk i\int 
                (J_{+}\varphi_{+}+K_{+}\chi_{+}+\eta_{+}\psi_{+}+
                    \bar\eta_{+} \bar\psi_{+}) 
               \rkk \rmk \rho\,\rkk \:, \non \\ 
\eea 

\noindent
where the suffix $c$ represents the closed time contour of integration
and $\varphi_{+}(\chi_{+},\psi_{+},\bar\psi_{+})$ a field component
$\varphi(\chi,\psi,\bar\psi)$ on the $+$-branch ($-\infty$ to $+\infty$),
$\varphi_{-}(\chi_{-},\psi_{-},\bar\psi_{-})$ that on the $-$-branch
($+\infty$ to $-\infty$). The symbol $T$ represents the time ordering
according to the closed time contour, $T_{+}$ the ordinary time
ordering, and $T_{-}$ the anti-time ordering. $J, K, \eta$, and
$\bar\eta$ imply the external fields for the scalar and the Dirac
fields, respectively. In fact, each external field
$J_{+}(K_{+},\eta_{+},\bar\eta_{+})$ and
$J_{-}(K_{-},\eta_{-},\bar\eta_{-})$ is identical, but for technical
reason we treat them different and set
$J_{+}=J_{-}(K_{+}=K_{-},\eta_{+}=\eta_{-},\bar\eta_{+}=\bar\eta_{-})$
only at the end of calculation. $\rho$ is the initial density
matrix. Strictly speaking, we need couple the time development of the
expectation value of the field with that of the density matrix, which
is practically impossible. Accordingly we assume that deviation from
the equilibrium is small and use the density matrix corresponding to
the finite \te state. Then the generating functional is described by
the path integral as
\beq
  Z[\,J,K,\eta,\bar\eta\,] 
    = \exp \biggl(\,iW[\,J,K,\eta,\bar\eta\,]\,\biggr) 
     = \int_{c}\CD\varphi \int_{c}\CD\chi \int_{c}\CD\psi 
        \int_{c}\CD\psi^{\ast}
        \exp{iS[\,\varphi,\chi,\psi,\bar\psi,J,K,\eta,\bar\eta\,]} \:,
\eeq 
where the classical action $S$ is given by
\beq
  S[\,\varphi,\chi,\psi,\bar\psi,J,K,\eta,\bar\eta\,]=
              \int_{c}d^4x \lhk \CL+J(x)\varphi(x)+K(x)\chi(x)
                 +\eta(x)\psi(x)+\bar\eta(x)\bar\psi(x) \rhk \:. 
\eeq 
As with the Euclidean-time formulation, the scalar field is still
periodic and the Dirac field anti-periodic along the imaginary
direction, now with $\varphi(t,\vect x)=\varphi(t-i\beta,\vect x)$, 
$\chi(t,\vect x)=\chi(t-i\beta,\vect x)$, and
$\psi(t,\vect x)=-\psi(t-i\beta,\vect x)$ \cite{per}.

The effective action for the scalar field is defined by the connected
generating functional as 
\beq
\Gamma[\phi]=W[\,J,K,\eta,\bar\eta\,]-\int_{c}d^4x J(x)\phi(x) \:,
   \label{eqn:a1effe}
\eeq
where $\phi(x)=\delta W[J,K,\eta,\bar\eta] / \delta J(x)$.

We give the finite \te propagator before the perturbative
expansion. For the closed path, the scalar propagator has four
components.
\bea G_{\chi}(x-x') &=& \left(
    \begin{array}{cc}
       G^{F}_{\chi}(x-x') & G^{+}_{\chi}(x-x')        \\
       G^{-}_{\chi}(x-x') & G^{\tilde F}_{\chi}(x-x') \\
    \end{array}
   \right) \non \\
                    &\equiv&
   \left(
    \begin{array}{cc}
      \mbox{Tr}[\,T_{+}\chi(x)\chi(x')\rho\,] & 
      \mbox{Tr}[\,\chi(x')\chi(x)\rho\,] \\
      \mbox{Tr}[\,\chi(x)\chi(x')\rho\,] & 
      \mbox{Tr}[\,T_{-}\chi(x)\chi(x')\rho\,] \\
    \end{array}
   \right) \non \\ &\equiv& 
   \int \frac{d^4 k}{(2\pi)^4}e^{-ik(x-x')}
   \left(
    \begin{array}{cc}
       G^{F}_{\chi}(k) & G^{+}_{\chi}(k)        \\
       G^{-}_{\chi}(k) & G^{\tilde F}_{\chi}(k) \\
    \end{array}
  \right) \:, 
\eea 
where
\bea
  G^{F}_{\chi}(k) &=& \frac{i}{k^2-m_{\chi}^2+i\epsilon}
                          +2\pi n_{\chi}(\vect k)   
                \,\delta(k^2-m_{\chi}^2) \:, \non \\
  G^{\tilde F}_{\chi}(k) &=& \frac{-i}{k^2-m_{\chi}^2-i\epsilon}
                                 +2\pi n_{\chi}(\vect k)   
                \,\delta(k^2-m_{\chi}^2) \:, \non \\
  G^{+}_{\chi}(k) &=& 2\pi\,[\,\theta(-k_{0})+n_{\chi}(\vect k)\,]   
                \,\delta(k^2-m_{\chi}^2) \:, \non \\
  G^{-}_{\chi}(k) &=& 2\pi\,[\,\theta(k_{0})+n_{\chi}(\vect k)\,]   
                \,\delta(k^2-m_{\chi}^2) \:,
\eea
with $n_{\chi}(\vect k)=(e^{\beta\omega_{\chi}(\vect k)}-1)^{-1}$,\,
$\omega_{\chi}(\vect k)^2=\vect k^2+m_{\chi}^2$,\, 
and $\epsilon(k_{0}) = \theta(k_{0})-\theta(-k_{0})$ \cite{pro}.
Similar formulae apply for $\varphi$ field as well.

Also, for a Dirac fermion we find
\bea S_{\psi}(x-x') &=& \left(
    \begin{array}{cc}
       S^{F}_{\psi}(x-x') & S^{+}_{\psi}(x-x')        \\
       S^{-}_{\psi}(x-x') & S^{\tilde F}_{\psi}(x-x') \\
    \end{array}
   \right) \non \\
                    &\equiv&
   \left(
    \begin{array}{cc}
      \mbox{Tr}[\,T_{+}\psi(x)\bar\psi(x')\rho\,] & 
      \mbox{Tr}[\,-\bar\psi(x')\psi(x)\rho\,] \\
      \mbox{Tr}[\,\psi(x)\bar\psi(x')\rho\,] & 
      \mbox{Tr}[\,T_{-}\psi(x)\bar\psi(x')\rho\,] \\
    \end{array}
   \right) \non \\  &\equiv& 
   \int \frac{d^4 k}{(2\pi)^4}e^{-ik(x-x')}
   \left(
    \begin{array}{cc}
       S^{F}_{\psi}(k) & S^{+}_{\psi}(k)        \\
       S^{-}_{\psi}(k) & S^{\tilde F}_{\psi}(k) \\
    \end{array}
  \right) \:, 
\eea 
where
\bea
  S^{F}_{\psi}(k) &=& \frac{i}{\not{k}-m_{\psi}+i\epsilon}
                       -2\pi n_{\psi}(\vect k)   
                (\not{k}+m_{\psi})\,\delta(k^2-m_{\psi}^2) \:, \non \\
  S^{\tilde F}_{\psi}(k) &=&  \frac{-i}{\not{k}-m_{\psi}-i\epsilon}
                               -2\pi n_{\psi}(\vect k)   
                (\not{k}+m_{\psi})\,\delta(k^2-m_{\psi}^2) \:, \non \\
  S^{+}_{\psi}(k) &=& 2\pi\, 
             [\,\theta(-k_{0})-n_{\psi}(\vect k)\,]\,(\not{k}+m_{\psi})
                \,\delta(k^2-m_{\psi}^2) \:, \non \\
  S^{-}_{\psi}(k) &=& 2\pi\,
             [\,\theta(k_{0})-n_{\psi}(\vect k)\,]\,(\not{k}+m_{\psi}) 
                \,\delta(k^2-m_{\psi}^2) \:,
\eea
with $n_{\psi}(\vect k)=(e^{\beta\omega_{\psi}(\vect k)}+1)^{-1},\, 
\omega_{\psi}(\vect k)^2=\vect k^2+m_{\psi}^2$ \cite{pro}.

\subsection{One-loop finite \te effective action}

\indent
The perturbative loop expansion for the effective action $\Gamma$ can
be obtained by transforming $\varphi \rightarrow \varphi_{0}+\zeta$
where $\varphi_{0}$ is the field configuration which extremizes the
classical action $S[\,\varphi,J\,]$ and $\zeta$ is small perturbation
around $\varphi_{0}$. Up to one loop order and $\CO(\lambda^2, g^4,
f^2)$, $\Gamma$ is made up of the graphs as depicted in Fig.\
\ref{fig:one}. Summing up these graphs, the effective action
$\Gamma$ becomes
\bea
  \Gamma[\phi_{c},\phi_{\Delta}] &= &\int d^4x \lhk
        \phi_{\Delta}(x)[\,-\Box-M^2-\tilde M^2\,]\phi_{c}(x)
        -\frac{\lambda}{4!}
           \lmk 
             4\phi_{\Delta}(x)\phi_{c}^3(x)+\phi_{c}(x)\phi_{\Delta}^3(x)  
           \rmk          
                                           \rhk \non \\
        &&-\int d^4x\int d^4x' 
           \lhk\, A_{1}(x-x') + A_{3}(x-x') \,\rhk
           \lkk\,
             \phi_{\Delta}(x)\phi_{c}(x)\phi_{c}^2(x')
             +\frac14 \phi_{\Delta}(x)\phi_{c}(x)\phi_{\Delta}^2(x')
           \,\rkk  \non \\
        &&-2\int d^4x\int d^4x' 
              A_{2}(x-x')\phi_{\Delta}(x)\phi_{c}(x') \non \\
        &&+\frac{i}{2}\int d^4x\int d^4x' 
          \biggl[\,
            \lhk\, B_{1}(x-x') + B_{3}(x-x') \,\rhk
            \phi_{\Delta}(x)\phi_{\Delta}(x')\phi_{c}(x)\phi_{c}(x')
                 \non \\  
         && \qquad \qquad \qquad \qquad
               +B_{2}(x-x')\phi_{\Delta}(x)\phi_{\Delta}(x')
          \,\,\biggr]  \:, \non \\
        &&
  \label{eqn:a3effe}
\eea
where
\bea
   \phi_{c} &\equiv& \frac12(\phi_{+}+\phi_{-}) \:, \\
   \phi_{\Delta} &\equiv& \phi_{+}-\phi_{-} \:, \\
   M^2 &=& m^2 + g^2 \int\frac{d^3\vect q}{(2\pi)^3}
             \frac{1+2n_{\chi}(\vect q)}{2\omega_{\chi}(\vect q)} 
                     \:, \\
   \tilde M^2 &=& \frac{\lambda}{2} \int\frac{d^3\vect q}{(2\pi)^3}
             \frac{1+2n_{\varphi}(\vect q)}
               {2\omega_{\varphi}(\vect q)} \:, \\
   A_{1}(x-x') &=& \frac{g^4}{2} 
                    \mbox{Im}\,[\,G^{F}_{\chi}(x-x')^2\,]
                      \,\theta(t-t')
         \label{eqn:A1}  \:. \\
   A_{2}(x-x') &=& f^2 \mbox{Im}\,[\,S_{\alpha\beta}^{F}(x-x')
                    S_{F}^{\beta\alpha}(x'-x)\,]\,\theta(t-t')
         \label{eqn:A2}  \:. \\
   A_{3}(x-x') &=& \frac{\lambda^2}{2} 
                    \mbox{Im}\,[\,G^{F}_{\varphi}(x-x')^2\,]
                      \,\theta(t-t')
         \label{eqn:A3}   \:. \\
   B_{1}(x-x') &=& \frac{g^4}{2} 
                    \mbox{Re}\,[\,G^{F}_{\chi}(x-x')^2\,] 
         \label{eqn:B1}  \:. \\
   B_{2}(x-x') &=& -f^2 \mbox{Re}\,[\,S_{\alpha\beta}^{F}(x-x')
                    S_{F}^{\beta\alpha}(x'-x)\,] 
         \label{eqn:B2}  \:. \\
   B_{3}(x-x') &=& \frac{\lambda^2}{2} 
                    \mbox{Re}\,[\,G^{F}_{\varphi}(x-x')^2\,]
         \label{eqn:B3}  \:,
\eea

The last term of (\ref{eqn:a3effe}) gives the imaginary contribution
to the effective action $\Gamma$. We can attribute these imaginary
terms to the functional integrals over Gaussian fluctuations $\xi_{1}$
and $\xi_{2}$ \cite{Mor}. That is to say, we can interpret that the
imaginary part of the effective action comes from random fluctuations
onto the expectation value.  Thus we rewrite (\ref{eqn:a3effe}) as
\beq
  \exp (i\Gamma[\phi_{c},\phi_{\Delta}])=\int\CD\xi_{1} \int\CD\xi_{2}
              P_{1}[\xi_{1}]P_{2}[\xi_{2}]\exp\lhk
                 iS_{\rm eff}[\,\phi_{c},\phi_{\Delta},\xi_{1},\xi_{2}\,]\rhk
                    \:,
\eeq
where
\beq
  S_{\rm eff}[\,\phi_{c},\phi_{\Delta},\xi_{1},\xi_{2}\,] \equiv
                 \mbox{Re}\Gamma
                  +\int d^4x[\,\xi_{1}(x)\phi_{c}(x)\phi_{\Delta}(x)
                    +\xi_{2}(x)\phi_{\Delta}(x)\,] \:,
 \label{eqn:a3cla}
\eeq
with the probability distribution functional
\beq
  P_{i}[\xi_{i}] = \CN_{i} \exp \lkk\, - \frac12\int d^4x \int d^4x'
                      \xi_{i}(x) \tilde B_{i}^{-1}(x-x')\xi_{i}(x')\,\rkk 
                       \qquad (i = 1, 2)    \:.
\eeq
where $\CN_{i}$ is a normalization factor and $\tilde B_{1}(x-x') =
B_{1}(x-x') + B_{3}(x-x'), \tilde B_{2}(x-x') = B_{2}(x-x')$.

\subsection{Equation of motion}
\label{subsec:lan}

\indent
Applying the variational principle to $S_{\rm eff}$, we obtain the
equation of motion for $\phi_{c}$.
\beq
  \frac{\delta S_{\rm eff}[\,\phi_{c},\phi_{\Delta},\xi_{1},\xi_{2}\,]}
        {\delta \phi_{\Delta}}  
   \biggl. \biggr|_{\phi_{\Delta}=0}       
      =0 \:.
\eeq

\noindent
>From (\ref{eqn:a3cla}) and (\ref{eqn:a3effe}), it reads
\bea
  (\,\Box&+&M^2+\tilde M^2\,)\,\phi_{c}(x)
          +\frac{\lambda}{3!}\phi_{c}^3(x) \non \\
      &+& \phi_{c}(x)\int d^3 \vect{x'}\int_{-\infty}^{t}dt'
              \tilde A_{1}(x-x')\phi_{c}^2(x')
           +2\int d^3
           \vect{x'}\int_{-\infty}^{t}dt'A_{2}(x-x')\phi_{c}(x')
             \non \\
           && \qquad \qquad \qquad \qquad \qquad \qquad \qquad
           =\phi_{c}(x)\xi_{1}(x)+\xi_{2}(x) \:,
  \label{eqn:a4eqm}
\eea
and
\beq
  \la\,\xi_{i}(x)\xi_{i}(x')\,\ra = \tilde B_{i}(x-x') \:,
  \label{eqn:corre}
\eeq

\noindent
where $\tilde A_{1}(x-x') = A_{1}(x-x') + A_{3}(x-x')$. Though $\tilde
A_{1}$ and $\tilde B_{1}$ has two contributions from $\chi$ and
$\varphi$ fields, they have the same properties except for the values
of coefficients and masses. From now on, we consider only the
contribution from $\chi$ field for simplicity and omit the suffix $c$. 
The right-hand-side of (\ref{eqn:a4eqm}) are the noise terms, while
the last two terms of the left-hand-side are combination of a
dissipation term and one-loop correction to the classical equation of
motion which would reduce to a derivative of the effective potential,
$V'_{\rm eff}(\phi)$, if we restricted $\phi(x')$ to be a constant in
space and time.

The above equation (\ref{eqn:a4eqm}) is an extension of equation (3.2)
of Gleiser and Ramos \cite{GR} in that we have incorporated not only
self-interaction but also interactions with a boson $\chi$ and a
fermion $\psi$. In \cite{GR} Gleiser and Ramos proposed to adopt
several further approximations to reduce their equation to the form of
the simple Langevin equation like (\ref{eqn:simplelangevin}) and
(\ref{eqn:whitenoise}). In particular, for the purpose of simplifying
the equation to a local form they handled spatial nonlocality by
considering only contributions with zero external momentum, which is
physically equivalent to dealing only with nearly spatially
homogeneous fields. With this approximation the correlation function
of the bosonic noise (\ref{eqn:B1}), for example, becomes
\bea
  \la\,\xi_{1}(x)\xi_{1}(x')\,\ra
         &\Rightarrow& \left.      
           \frac{g^4}{2}
            \int\frac{d^3\vect k}{(2\pi)^3}
             e^{i\vect k\cdot(\vect x-\vect x')}
              \int\frac{d^3\vect q}{(2\pi)^3}
                \mbox{Re}\,[\,G^{F}_{\chi}(\vect q,t-t')
                   G^{F}_{\chi}(\vect{q-k},t-t')\,] 
                    \right|_{\vect{k}=\vect{0}}
                      \:, \non  \\
         &=& \frac{g^4}{2} \delta^3(\vect x-\vect x')
               \int\frac{d^3\vect q}{(2\pi)^3}
                \mbox{Re}\,[\,G^{F}_{\chi}(\vect q,t-t')\,]^2
                 \:. 
             \label{eqn:homas} 
\eea

\noindent
We thus obtain spatially uncorrelated noise, which would violate
spatial homogeneity of $\phi$ in the severest manner and lead to a
self-inconsistent result.

Since the noise term in (\ref{eqn:a4eqm}) are the only source of
inhomogeneous evolution of $\phi$, we should not adopt the above
approximation (\ref{eqn:homas}) for their correlations. Instead,
keeping the original form of the correlation functions of the noises,
(\ref{eqn:corre}) with (\ref{eqn:B1}) through (\ref{eqn:B3}), we can
obtain important informations about inhomogeneous fluctuations which
help us to set the lattice spacing of simulations. For this purpose we
calculate the spatial correlation functions of the noises more
explicitly in the next subsection.

\subsection{Spatial correlation of noises}

\indent
>From the above discussion, we see the correlation length of the noise
is the most important scale in order to investigate the effect of
fluctuations on the dynamics of the phase transition.  The
correlation function of the noises are given in (\ref{eqn:corre}), which are,
unfortunately, too complicated to apply for numerical simulations
directly. Since nontrivial temporal correlation is expected to affect
the relaxation process and its time scale only, we adopt an
approximation that temporal correlation is white but take spatial
correlation fully into account. So we evaluate the equal-time spatial
correlation.

First we consider the case of the bosonic noise.  The equal-time
propagator in momentum space is given by \cite{pro}
\bea 
  G^{F}_{\chi}(\vect k,0) &=& \frac{1}{2\omega_{k}}
          \,\biggl[\,n_{\chi}(-\omega_{k})
           +n_{\chi}(-\omega_{k})-1 \,\biggr] \non \\ 
                   &=&
             \frac{1}{2\omega_{k}}
              \lkk\,1+2\sum_{n=1}^{\infty}e^{-n\beta
               \omega_{k}} \,\rkk \:.
\eea 
Then that in configuration space propagator reads
\bea
  G^{F}_{\chi}(\vect x,0) &=& \int \frac{d^3\vect k}{(2\pi)^3}
                    e^{i\vect k \cdot \vect x} G^{F}_{\chi}(\vect k,0)
                        \non  \\
                   &=& \int \frac{d^3\vect k}{(2\pi)^3} 
                         \frac{e^{i\vect k \cdot \vect x}}{2 \omega_{k}} 
                          \lkk\,1+2\sum_{n=1}^{\infty}e^{-n\beta 
                            \omega_{k}}
                          \,\rkk   \non \\
                   &=& \frac{m_{\chi}}{4\pi^{2}r}
                        \lkk\,K_{1}(m_{\chi}r)+2\sum_{n=1}^{\infty}
                          \frac{r}{\sqrt{r^2+n^2 \beta^2}}
                           \,K_{1}(m_{\chi}\sqrt{r^2+n^2 \beta^2}\,)
                        \,\rkk, \quad r = |\vect x| \:,
\eea
which of course has the same form as (\ref{eqn:nonself}).
Using the above representation, the equal-time spatial correlation of
the bosonic noise becomes
\bea
  \la\,\xi_{1}(\vect x)\xi_{1}(\vect 0)\,\ra_{\rm{equal-time}}
             &=& \frac{g^4}{2}\,\mbox{Re} \biggl[ 
                    G^{F}_{\chi}(\vect x,0)^2  
                        \,\biggr] \non \\
             && \hspace{-3.3cm}
                  =\frac{m_{\chi}^2 g^4}{32\pi^4 r^2} 
                    \lkk\,K_{1}(m_{\chi}r)+2\sum_{n=1}^{\infty}
                          \frac{r}{\sqrt{r^2+n^2 \beta^2}}
                           \,K_{1}(m_{\chi}\sqrt{r^2+n^2 \beta^2}\,)
                    \,\rkk^2 \;,
\eea
which damps exponentially for $r > m_{\chi}^{-1}\:.$

For the massless bosonic noise, we can calculate the sum of the
infinite series to yield
\beq
  G^{F}_{\chi}(\vect x,0) = \frac{1}{4\pi \beta r}
                      \coth \lmk \frac{r}{\beta}\pi \rmk \:.
\eeq
and the equal-time spatial correlation of the bosonic noise becomes
\bea
  \la\,\xi_{1}(\vect x)\xi_{1}(\vect 0)\,\ra_{\rm{equal-time}}
            &=& \frac{g^4}{32 \pi^2 \beta^2 r^2}
                 \coth^2 \lmk \frac{r}{\beta}\pi \rmk \:, \\
            &\simeq& 
                  \frac{g^4}{32 \pi^2 \beta^2}
                   \frac{1}{r^2}
                       \qquad \mbox{for}\:, r \gg \frac{\beta}{\pi} \:.
\eea
We thus find that for the massive bosonic noise the correlation function
damps exponentially, particularly the damping scale is the inverse of
mass, while for the massless bosonic noise it damps
much less rapidly, according to a power-law.

For the fermionic noise, similarly, the equal-time propagator in
momentum space is given by \cite{pro}
\bea
  S^{F}_{\psi}(\vect k,0) &=& 
         \frac{1}{2\omega_{k}}(m_{\psi}-\vect\gamma \cdot
         \vect k) \,\biggl[\,n_{\psi}(-\omega_{k})
           +n_{\psi}(-\omega_{k})-1 \,\biggr]  \non  \\ 
                   &=&
             \frac{1}{2\omega_{k}}(m_{\psi}-\vect \gamma \cdot \vect k)
              \lkk\,1+2\sum_{n=1}^{\infty}(-1)^{n}e^{-n\beta
               \omega_{k}} \,\rkk \:, 
\eea
and that in configuration space reads
\bea
  S^{F}_{\psi}(\vect x,0) &=& \int \frac{d^3\vect k}{(2\pi)^3}
                    e^{i\vect k \cdot \vect x} S^{F}_{\psi}(\vect k,0)
                        \non   \\
                   &=& (m_{\psi}+i\vect\gamma \cdot \vect\nabla)
                        \int \frac{d^3\vect k}{(2\pi)^3} 
                         \frac{e^{i\vect k \cdot \vect x}}{2 \omega_{k}} 
                          \lkk\,1+2\sum_{n=1}^{\infty}(-1)^{n}e^{-n\beta 
                            \omega_{k}}
                          \,\rkk  \non \\
                   &=& \frac{m_{\psi}^{2}}{4\pi^{2}r}
                        \lkk\,K_{1}(m_{\psi}r)+2\sum_{n=1}^{\infty}(-1)^n
                          \frac{r}{\sqrt{r^2+n^2 \beta^2}}
                           \,K_{1}(m_{\psi}\sqrt{r^2+n^2 \beta^2}\,)
                        \,\rkk  \non \\
                   && \hspace{-0.5cm} -\,\frac{im_{\psi}^2}{4\pi^2 r}
                        \frac{\vect\gamma \cdot \vect x}{r}
                         \lkk\,K_{2}(m_{\psi}r)+2\sum_{n=1}^{\infty}(-1)^n
                           \frac{r^2}{r^2+n^2 \beta^2}
                            \,K_{2}(m_{\psi}\sqrt{r^2+n^2 \beta^2}\,)
                         \,\rkk \:. 
\eea
Also, the equal-time correlation of the fermionic noise is given by
\bea
  \la\,\xi_{2}(\vect x)\xi_{2}(\vect 0)\,\ra_{\rm{equal-time}}
             &=& -f^2\,\mbox{Re} \biggl[\,\mbox{Tr} 
                   \biggl(S^{F}_{\psi}(\vect x,0)
                    S^{F}_{\psi}(-\vect x,0) \biggr)  
                        \,\biggr] \non \\
             && \hspace{-3.3cm}
                  =\frac{m_{\psi}^4 f^2}{4\pi^4 r^2} \lhk 
                    \lkk\,K_{2}(m_{\psi}r)+2\sum_{n=1}^{\infty}(-1)^n
                          \frac{r^2}{\sqrt{r^2+n^2 \beta^2}}
                           \,K_{2}(m_{\psi}\sqrt{r^2+n^2 \beta^2}\,)
                    \,\rkk^2 \right. \non \\ 
             && \hspace{-3.0cm}    \left.
                   -\lkk\,K_{1}(m_{\psi}r)+2\sum_{n=1}^{\infty}(-1)^n
                          \frac{r}{r^2+n^2 \beta^2}
                           \,K_{1}(m_{\psi}\sqrt{r^2+n^2 \beta^2}\,)
                    \,\rkk^2 \,\rhk \:,     
\eea
which damps exponentially at the inverse mass scale for $m_{\psi}\beta
\gg 1$, and at the scale $\beta$ for $m_{\psi}\beta \ltilde 1$.
For the massless fermion, we find
\beq
  S^{F}_{\psi}(\vect x,0) = 
         -\frac{i}{4\pi^2 r^2}\,\vect\gamma \cdot \vect x
          \lkk\, \frac{\pi}{\beta r}
                  \frac{1}{\sinh \lmk \frac{r}{\beta}\pi \rmk} 
                +\frac{\pi^2}{\beta^2}
                  \frac{\cosh \lmk\frac{r}{\beta}\pi \rmk}   
                       {\sinh^2 \lmk \frac{r}{\beta}\pi \rmk}
          \,\rkk \:.  
\eeq
and
\bea
  \la\,\xi_{2}(\vect x)\xi_{2}(\vect 0)\,\ra_{\rm{equal-time}}
            &=& \frac{f^2}{4 \pi^4 r^2}
                  \lkk\, \frac{\pi}{\beta r}
                   \frac{1}{\sinh \lmk \frac{r}{\beta}\pi \rmk} 
                 +\frac{\pi^2}{\beta^2}
                   \frac{\cosh \lmk\frac{r}{\beta}\pi \rmk}   
                        {\sinh^2 \lmk \frac{r}{\beta}\pi \rmk}
                  \,\rkk^2  \:, \\
            &\simeq& 
                  \frac{f^2}{4\beta^4}
                    \frac{e^{-\frac{2\pi r}{\beta}}}{r^2} \;. 
                       \qquad \mbox{for}\:, r \gg \frac{\beta}{\pi} \:.   
\eea

Note that for the fermionic noise, unlike the bosonic case, the
correlation function damps exponentially regardless of the mass.  This
is the very interesting feature of the noise. We can physically
interpret this feature as follows. Since there is Pauli's blocking law
for a fermion, particles are apt to separate from one another and the
correlation is easily destructed. On the other hand, bosonic particles
can occupy the same state and the correlation is kept. When the \te is
zero, both fermionic and bosonic massless propagators damp according
to a power-law.  The above feature is an example of the fact that
statistical difference between fermions and bosons appear more
markedly at finite \te than at zero one.

\subsection{Dissipation term}

\indent

The equation of motion (\ref{eqn:a4eqm}) derived in the subsection
\ref{subsec:lan} has contributions representing the dissipative
effect in the last two terms of the left hand side. Since these terms
are nonlocal in time, what is often done in the literatures
\cite{Mor} \cite{GR} to extract local terms proportional to
$\dot\phi$ is to assume that the field changes adiabatically, or
put

\beq
  \phi^n(\vect x',t') \simeq \phi^n(\vect x',t)
   - n(t'-t) \phi^{n-1}(\vect x',t) \dot\phi(\vect x',t) \:,
  \label{eqn:adia}
\eeq 

\noindent
in the integrand of (\ref{eqn:a4eqm}). But the dissipation terms thus
evaluated vanish as long as we use the bare propagators. This is
usually interpreted as a manifestation of the fact that the
dissipative effect is intrinsically a non-perturbative phenomenon and
cannot be investigated from the perturbation theory \cite{BVHLS}. In
order to see a damping effect, we must observe the system for a finite
duration of time typically proportional to the inverse of some
coupling constants. After this period, however, the perturbation
theory may have broken down as explained in \cite{BVHLS} using a toy
model. So, in order to find the damping effect, non-perturbative terms
are often incorporated by using a ``dressed'' propagator with an
explicit width, which is obtained by resuming higher-loop graphs of
some classes, instead of the bare one \cite{Mor,GR}. 
Although we may obtain finite
dissipation terms in this way, they do not yet satisfy the
fluctuation-dissipation relation. More serious is the fact that these
approach is not self-consistent as criticized by Gleiner and M\"uller
\cite{GM}. In fact, apart from the validity of perturbation, the
adiabatic expansion (\ref{eqn:adia}) itself breaks down before we can
observe dissipation effect \cite{GM}.

In order to cure the situation, Gleiner and M\"uller \cite{GM} proposed
to adopt the linear harmonic expansion of the Fourier mode $\phi(\vect
k,t')$ as
\beq
  \phi(\vect k,t') \simeq \phi(\vect k,t)
                            \cos \lkk\,\omega_{k}(t'-t)\,\rkk
             - \dot\phi(\vect k,t) \frac{1}{\omega_{k}}
                            \sin \lkk\,\omega_{k}(t'-t)\,\rkk
\eeq
and calculated the dissipation term in a simple model. They have shown
that the fluctuation-dissipation relation is met in the classical
limit \cite{GM}. Unfortunately, however, the applicability of this
method is rather limited and we cannot calculate the dissipation term
in our model explicitly. Therefore, here we make much of the
thermodynamics and derive it so that the fluctuation-dissipation
relation is met. We also identify the remaining part of the integral
terms of (\ref{eqn:a4eqm}) with the derivative of the effective
potential $V'_{\rm eff}(\phi)$ , primarily for simplicity. But if the
system turns out to be homogeneous as a result of numerical
simulations, this choice will be justified since the homogeneous
expectation value should take a value where the effective potential is
minimized, but otherwise unjustified. In the latter case the one-loop
approximation also breaks down and then we would have to deal with the
full effective action, which is beyond the scope of the present
analysis. We thus interpret the integral terms in (\ref{eqn:a4eqm}) as
consisting of two parts, one contribution to the derivative of the
effective potential and the other to the dissipation term, and this
division is done so that the dissipation term satisfies the
fluctuation-dissipation relation. Note that, although the above
procedure is physically motivated, it should be regarded as an {\it
 ansatz} rather than an approximation and we have been unable to
derive these terms rigorously from first principles at present.
Nevertheless we stress that our approach is self-consistent if the
field configuration turns out to be homogeneous. 

We next put the above scheme into practice. First we consider the case 
thermalization proceeds only through bosonic noises and determine the
corresponding dissipation term. To do this we rewrite
(\ref{eqn:a4eqm}) in the form

\beq 
  \Box \phi(x) + V'_{\rm eff}(\phi) 
    +\phi(x)\int d^3 \vect x'\CA_{1}(\vect x-\vect x',t)
      \phi(\vect x',t)\dot\phi(\vect x',t)
        = \phi(x)\xi_{1}(x) \:,
\eeq 

\noindent
where $\CA_{1}(\vect x-\vect x',t)$ is to be determined so that the
fluctuation-dissipation relation is met. For the purpose of applying
to numerical simulations we adopt an approximation that $\xi_{1}(x)$
is a temporally white noise with $\la\,\xi_{1}(x)\xi_{1}(x)\,\ra =
\CB_{1}(\vect x-\vect x')\delta(t-t')$, which is a good approximation
since temporal correlation damps exponentially beyond
$|t-t'|>\beta/(2\pi)$. 
Then this Langevin equation
can be converted into the Fokker-Plank equation.

\bea
  \frac{\del W}{\del t} &=& \int d^3\vect x \lhk
      -\frac{\delta\CH}{\delta\pi(x)}\frac{\delta W}{\delta\phi(x)}
       +\frac{\delta\CH}{\delta\phi(x)}\frac{\delta W}{\delta\pi(x)}
        \right. \non \\*    
         && \left. \qquad \quad  
        +\,\frac{\delta}{\delta\pi(x)}
         \biggl[\,
            \phi(\vect x,t)\int d^3 \vect x'\CA_{1}(\vect x-\vect x',t)
              \phi(\vect x',t)\pi(\vect x',t)\cdot W   
         \biggr] \rhk \non \\* 
           && +\,\frac12\int d^3\vect x \int d^3\vect x'
               \phi(\vect x,t)\phi(\vect x',t)
                \CB_{1}(\vect x-\vect x')\frac{\delta^2 W}
                 {\delta\pi(x) \delta\pi(x')} \:,
\eea

\noindent
where $W[\phi(\vect x,t),\pi(\vect x,t)]$ is the distribution
function, $\pi(x)=\dot\phi(x)$, and $\CH$ is the Hamiltonian, which is
given by $\CH = \int d^3\vect x \lkk\,\frac12\,\pi^2 +
\frac12(\vect\nabla\phi)^2+V_{\rm eff}(\phi)\,\rkk$. In order that
this equation has a stationary solution, it is at least necessary that
$\CA_{1}(\vect x-\vect x',t)$ does not depend on time, namely,
$\CA_{1}(\vect x-\vect x',t)=\CA_{1}(\vect x-\vect x') \:.$ We require
that $W_{\rm st} \equiv \CN e^{-\frac{\CH}{T}}$ constitutes a stationary
solution of the above equation,

\beq
  0 = \int d^3\vect x \int d^3\vect x'\phi(\vect x,t)\phi(\vect x',t)
       \lkk\,
         \delta^3(\vect x-\vect x')W-\frac1T\pi(\vect x)\pi(\vect x')W
       \,\rkk
        \lkk\,
          \CA_{1}(\vect x-\vect x')-\frac{\CB_{1}(\vect x-\vect x')}{2T}
        \,\rkk \:. 
\eeq
Then we find
\beq
  \CA_{1}(\vect x-\vect x') = \frac{1}{2T}\CB_{1}(\vect x-\vect x') \:,
\eeq
and 
\beq
  \Box \phi(x) + V'_{\rm eff}(\phi)
     +\frac{1}{2T}\phi(x) \int d^3 \vect x'\CB_{1}(\vect x-\vect x')
                      \phi(\vect x',t)\dot\phi(\vect x',t) \non \\
            = \phi(x)\xi_{1}(x) \:.
\eeq

The fermionic contribution can be treated similarly, and we find
\beq
  \Box \phi(x) + V'_{\rm eff}(\phi)
        + \frac{1}{2T}\int d^3 \vect x'
               \CB_{2}(\vect x-\vect x')\dot\phi(\vect x',t) 
            = \xi_{2}(x) \:,
\eeq   
\beq
  \la\,\xi_{2}(x)\xi_{2}(x)\,\ra = 
     \CB_{2}(\vect x-\vect x')\delta(t-t') \:,
\eeq
in the case thermalization is realized only through fermionic noises.

\section{Numerical simulations}

\label{sec:numerical}

\indent

Using the phenomenological Langevin equation obtained in the previous
section, we now perform numerical simulations in the same way as in
Sec.\ \ref{sec:re-analysis}. The essential difference from Borrill and
Gleiser's formulation \cite{BG} is that we generate noises which have
spatial correlations calculated in the previous section.

In order to approach some aspects of the electroweak phase transition
in the standard model, we adopt the one-loop improved effective
potential of the Higgs field in our phenomenological Langevin
equation. Since the properties of massless bosonic noises such as
those from gauge interactions and fermionic noises are very different
from each other, we analyze the cases thermalization proceed through
bosonic noises and fermionic noises separately.

\subsection{Bosonic noise}

\indent
We first examine an extreme case that thermalization of $\phi$
proceeds by virtue of only massless bosonic noises which has a
power-law correlation function such as those arising from gauge
interactions.  As a power-law is scale free, we set the same lattice
spacing as the fermionic case, $\beta/(2\pi)$, in order to see the
difference between the behavior of bosonic noises and that of
fermionic counterparts.

Before calculation it should be noted that the bosonic noise we have
derived so far is multiplicative and accordingly, if we set the
initial condition as $\Phi=\dot\Phi=0$, the system does not
evolve. Hence, in this case we add the following contributions with
two loops and order $g^4$ (Fig.\ \ref{fig:two}), which lead the
additive noise. Then the correction to the effective potential is
given by
\bea
  \Delta\Gamma[\phi_{c},\phi_{\Delta}] &=& - \int d^4x
          \phi_{\Delta}(x)\Delta V \phi_{c}(x)    
       -2\,\int d^4x\int d^4x' A_{4}(x-x')\phi_{\Delta}(x)\phi_{c}(x')
              \non \\ 
       && +\frac{i}{2}\int d^4x\int d^4x' 
           B_{4}(x-x')\phi_{\Delta}(x)\phi_{\Delta}(x') \:,
\eea
where
\bea
   \Delta V   &=& \frac{g^4}{2} \int dt'\int\frac{d^3\vect k}{(2\pi)^3}
                \mbox{Im} \lkk\,G^{F}_{\chi}(\vect k,t-t')
                 \,\rkk^2 \theta(t-t')
                  \int\frac{d^3\vect q}{(2\pi)^3} 
                   \frac{1+2n_{\varphi}(\vect q)}
                        {2\omega_{\varphi}(\vect q)} \:, \\           
   A_{4}(x-x') &=& \frac{g^4}{2} 
                    \mbox{Im}\,
                [\,G^{F}_{\chi}(x-x')^2\,G^{F}_{\varphi}(x-x')\,]
                      \,\theta(t-t') \:, \\
   B_{4}(x-x') &=& \frac{g^4}{2} 
                    \mbox{Re}\,
                [\,G^{F}_{\chi}(x-x')^2\,G^{F}_{\varphi}(x-x')\,] \;.
\eea
The last term is imaginary and we regard it as the term coming from
the stochastic noise as before. The dissipation term is obtained so
that the fluctuation-dissipation relation is met. Then the
dimensionless equation of motion only with bosonic contributions
becomes,
\bea
  \frac{\del^2 \Phi}{\del t^2}(x) &=& \nabla^2 \Phi(x)
             - \frac{\del U(\Phi)}{\del \Phi} \non \\
     && -\frac{1}{2\theta}\Phi(x) 
         \int d^3 {\vect x'}
          \CB_{1}({\vect x}-{\vect x'})
           \Phi({\vect x'},t)\dot \Phi({\vect x'},t)
        -\frac{1}{2\theta}\int d^3 {\vect x'}
         \CB_{4}({\vect x}-{\vect x'})
          \dot \Phi({\vect x'},t) \non \\
       && \qquad \qquad \qquad \qquad \qquad \qquad \qquad 
        + \Phi(x)\xi_{1}(x)+\xi_{4}(x) \:,
\eea
where
\bea
  \la\,\xi_{1}(x)\xi_{1}(x')\,\ra &=& 
    \CB_{1}({\vect x}-{\vect x'})\delta(t-t') =
       \frac{1}{|\vect x-\vect x'|^2}\delta(t-t') \:, \\
  \la\,\xi_{4}(x)\xi_{4}(x')\,\ra &=& 
    \CB_{4}({\vect x}-{\vect x'})\delta(t-t') =
      \frac{\theta_{c}}{4\pi}
       \frac{1}{|\vect x-\vect x'|^3}\delta(t-t') \;.
\eea  
In reality, the correlation function of the noises as derived from a
fundamental theory is proportional to some powers of coupling
constants. However, once the fluctuation-dissipation relation is
assumed, its magnitude does not affect the final equilibrium
configuration at all. Hence we have normalized the amplitude of noises
as above in our numerical calculations.

Noises with the above correlation can easily be generated working in
Fourier space. Since we are assuming $\xi(\vect x)$ is random
Gaussian, its Fourier transform,
\beq 
  \zeta(\vect k) = 
       \int d^3 \vect x\,e^{-i \vect k\cdot\vect x}\xi(\vect x)
     \:,
\eeq 
also satisfies the Gaussian probability function, and the distribution
function for each mode becomes
\beq
  P[\,\zeta(\vect k)\,]=N'\exp\biggl[ -\frac12 \int 
            \frac{d^3\vect k}{(2\pi)^3}
             \frac{|\zeta(\vect k)|^2}{P(\vect k)}\,\biggr] \:.
\eeq
Here $N'$ is the normalization factor, and $P(\vect k)$ is the power
spectrum given by
\beq
  P(\vect k)=\int d^3 \vect x\,e^{-i\vect k\cdot\vect x}\CB(\vect x)
    \:,
\eeq
and we have used the reality condition for $\xi(\vect x)$. This
distribution has no correlation between each Fourier mode. Therefore
we have only to generate the white noise in the momentum space and
Fourier transform it into the configuration space.

Next we discretize the system. Then the discretized master equation
becomes 
\bea 
  \dot\Phi_{\vect i,n+1/2} &=& 
       \Biggl[\,
          \lmk\,
             1-\frac{\delta t \delta x}{4\theta}\Phi_{\vect i,n}
              \Phi_{\vect i,n}B_{1\,\vect 0} 
               - \frac{1}{4\theta}B_{4\,\vect 0}
          ,\rmk \dot\Phi_{\vect i,n-1/2} 
       \Biggr. \non \\
             && \, 
           +\,\delta t 
          \lmk\,
             \nabla^2\Phi_{\vect i,n}
             -\left. \frac{\del U}{\del\Phi}\right|_{\vect i, n}  
             -\frac{\delta x}{2\theta}\Phi_{\vect i,n} \sum_{\vect j}
                  (1-\delta_{\vect i \vect j}) B_{1\,\vect i-\vect j}
                  \Phi_{\vect j,n} \dot\Phi_{\vect j,n-1/2} 
           \right.  \non \\  
             && \,\,
       \left. 
          \left.    
         -\frac{1}{2\theta} \sum_{\vect j} (1-\delta_{\vect i \vect j})
                  B_{4\,\vect i-\vect j} \dot\Phi_{\vect j,n-1/2} 
             + \Phi_{\vect i,n} \xi_{1\,\vect i,n}
             + \xi_{2\,\vect i,n}
          \,\rmk  
       \,\rkk  \non \\
       && \times
          \lmk
            1+\frac{\delta t \delta x}{4\theta}\Phi_{\vect i,n}
            \Phi_{\vect i,n}B_{1\,\vect 0}+\frac{1}{4\theta}B_{4\,\vect 0}
          \rmk^{-1}
   \:, \non \\
  \Phi_{\vect i,n+1} &=& \Phi_{\vect i,n} + \delta t\,\dot\Phi_{\vect
    i,n+1/2} 
   \:, \non \\          
   \nabla^2\Phi_{\vect i,n} &\equiv&  \sum_{s = x, y, z} 
       \frac{\Phi_{i_{s}+1_{s},n} - 2\,\Phi_{i_{s},n} +
             \Phi_{i_{s}-1_{s},n}}{(\delta x)^2} \:,
\eea     
where we have used the second-order leapfrog method and adopted the
Crank-Nicholson scheme only for the diagonal term because of its
dominance for numerical stability. We also set $B_{1\,\vect i} = 1 / |\vect
i|^2, B_{4\,\vect i} = 1 / |\vect i|^3 $ except $B_{1\,\vect 0} = 1 /
N^2, B_{4\,\vect 0} = 1 / N^3 $. Initial conditions are given by
$\Phi_{\vect i,0} = \dot\Phi_{\vect i,0} = 0$ for all $\vect i$'s.

The results are depicted in Fig.\ \ref{fig:boson}. Massless bosonic
noises do not disturb the homogeneous field configuration at least for 
small enough values of $\lambda$.

\subsection{Fermionic noise}

\label{subsec:fermion}

\indent

Next we consider the fermionic case.  The spatial correlation of the
fermionic noise damps exponentially for both massive and massless
fermions. The fermions interacting with the Higgs field in the standard
model acquire finite masses if and only if the Higgs field has a
non-vanishing expectation value. Hence in the same spirit as adopting
the effective potential in the equation of motion of the field, we
perform numerical calculations with massless fermionic noises. That
is, if $\phi$ remains to be equal to zero homogeneously reflecting the
initial condition, this choice is consistent.  In fact, however, even
when phase mixing is manifest, the expectation value of the Higgs
field remains at most about 50GeV for $M_{H} \simeq 60$GeV, which
means that quark masses remain much smaller than the \te and massless
approximation itself is justified in this case as well.

Since the correlation of the fermionic noise damps exponentially, we
can use the white noise as long as we take the lattice spacing larger
than the correlation length, which is equal to $\beta/(2\pi)$ and in
our dimensionless unit it corresponds to $\delta x =
(2D)^{1/2}/(2\pi \theta_{c})$ (= 0.092 for $M_{H} = 60$GeV). Then the
master equation is the same as that used by Borrill and Gleiser
\cite{BG} or equation (\ref{eqn:lesseq}),

\beq
  \frac{\del^2 \Phi}{\del t^2}=\nabla^2 \Phi-\eta \frac{\del \Phi}{\del t} 
     -\frac{\del U(\Phi)}{\del \Phi}+\xi_{2}(x) \:,
\eeq
where
\beq
  \la\,\xi_{2}(x)\xi_{2}(x')\,\ra
       = 2\eta\theta\delta(t-t')\delta^3(\vect x-\vect x') \:.
\eeq

The results of the solutions of this master equation has already been
depicted in Fig.\ \ref{fig:lambda} for various grid spacing $\delta
x$. Figure \ref{fig:fermion} represents the case $\delta x$ is set to
be the fundamental length. As is seen in these figures, the fermionic
noises are more effective to disturb the field configuration from a
homogeneous state to an inhomogeneous one with possible mixing of two
phases.

\section{Analytic interpretation of numerical results}

\label{sec:analytic}

\indent

As in Fig.\ \ref{fig:lambda}, in some choice of $\delta x$ and
$\lambda$ the system seems to relax to a state with $f_{0}$ some value
between $f_{0} = 0.5$ and $1$. One may wonder that this is because our
simulation time is so short that the system still keeps the memory of
the particular initial condition, and that if we could trace the
evolution for a long enough period the system would relax into a state
with $f_{0} = 0.5$. In fact, however, if we examine the time variation
of the field configuration by observing its snapshot at different
times as depicted in Figs. 12(a) - (e), we can easily convince
ourselves that the system is in a stationary state with a constant
$f_{0} (> 0.5)$ repeating creation and annihilation of a number of
small domains of the asymmetric state whose typical radius is at most
a few times the lattice spacing. In this section we would like to
present an analytic argument to support that the state we followed in
the numerical simulation is a thermal state. Similar analysis has
already been done by Gleiser, Heckler, and Kolb \cite{GHK} in a
slightly different situation. See also Gelmini and Gleiser \cite{GG}.

Let $d_{+}(R,\Phi,t)$ be the number density of the asymmetric-state
bubbles with radius $R ( > \delta x)$ and amplitude $\Phi ( >
\Phi_{-})$ at $t$. In order to obtain the Boltzmann equation for
$d_{+}(R,\Phi,t)$, we count the processes which change it: i)\ Thermal
nucleation of asymmetric-state bubbles in an almost homogeneous
symmetric-state sea. ii)\ Annihilation of these asymmetric-state
bubbles into the symmetric-state sea. This is to be distinguished from the
process of nucleation of a symmetric-state bubble in a homogeneous
background of the asymmetric-state whose rate would be identical to
that of i) in a degenerate potential. We expect the process i) has
smaller rate than the process ii), since the former requires more
energy. We do not take into account the process that nucleated bubbles
dynamically shrink ($|v|\del d_{+}/\del R$ term in \cite{GHK}) because
the typical radius of nucleated bubbles is comparable to the lattice
spacing and shrinking bubbles and vanishing bubbles are hardly
distinguishable in our simulations. Then the Boltzmann equation for
$d_{+}(R,\Phi,t)$ becomes

\beq
  \frac{\del d_{+}(R,\Phi,t)}{\del t} = 
       (1 - f_{+}(t))\,G_{(\rm{s-phase} \Rightarrow R,\Phi)}
     - \frac{4 \pi R^3}{3} d_{+}(R,\Phi,t)\, 
          G_{(\rm{s-phase} \Leftarrow R,\Phi)} \:,
\eeq

\noindent
where $G_{(\rm{s-phase} \Rightarrow R,\Phi)}$ is the nucleation rate
for the process i), $G_{(\rm{s-phase} \Leftarrow R,\Phi)}$ is that for
the process ii).  We also assume that nucleation rates can be obtained
from the Gibbs distribution, namely $G = A \exp(-F/\theta_{c})$, where
$A$ is a constant. For $G_{(\rm{s-phase} \Rightarrow R,\Phi)}$ we put
$F = b\Phi^2 R$ taking surface tension of the created bubbles into
account with $b$ a constant. Since the inverse process is not
Boltzmann suppressed, we assume its rate is a constant:
$G_{(\rm{s-phase} \Leftarrow R,\Phi)} = B$.

In an equilibrium state, $\del d_{+}/\del t$ equals zero for all
$\Phi$'s and $R$'s,
\beq
  (1 - f_{+}^{eq})\,\,
     \frac{G_{(\rm{s-phase} \Rightarrow R,\Phi)}}
          {G_{(\rm{s-phase} \Leftarrow R,\Phi)}}     
        = \frac{4 \pi R^3}{3} d_{+}(R,\Phi,t) \:.
\eeq
Summing these equations for all $\Phi$'s and $R$'s leads to
\beq
  \int_{\delta x}^{\infty} \int_{\Phi_{-}}^{\infty}
    (1 - f_{+}^{eq})\,\,
     \frac{G_{(\rm{s-phase} \Rightarrow R,\Phi)}}
          {G_{(\rm{s-phase} \Leftarrow R,\Phi)}}
       \,d\Phi dR  
     = \int_{\delta x}^{\infty} \int_{\Phi_{-}}^{\infty}
         \frac{4 \pi R^3}{3} d_{+}(R,\Phi,t) \,d\Phi dR  
     = f_{+}^{eq} \:.
\eeq
Then $f_{+}^{eq}$ becomes
\beq 
  f_{+}^{eq} = \frac{I}{I + 1} \:,
\eeq
where
\bea
  I &=& \int_{\delta x}^{\infty} \int_{\Phi_{-}}^{\infty}\,
     \frac{G_{(\rm{s-phase} \Rightarrow R,\Phi)}}
          {G_{(\rm{s-phase} \Leftarrow R,\Phi)}}
        \,d\Phi dR       \non \\
    &=& \frac{A}{B}\,\int_{\delta x}^{\infty} \int_{\Phi_{-}}^{\infty}
             \exp(-F/\theta_{c}) \,d\Phi dR   \:.
\eea
$f_{+}^{eq}$ is depicted in Fig.\ \ref{fig:analytical} as a function
of $\lambda$. Since the case for $\delta x = 1.0$ is suitable for
seeing the change of the asymmetric fraction from non-mixing to
percolation, we set $\delta x = 1.0$. Then we can fit the analytic
solution with the numerical simulation very well with $A/B = 65$ and
$b = 2.77$.

Note that even if $|v|$ equals zero, $f_{+}^{eq}$ can become values
different from 0.5. The essence lies in the fact that creation rate of
an asymmetric-state domain in the symmetric phase and that of its
reverse can be different due to the surface tension even at the
critical temperature if the background is sufficiently homogeneous.
Of course, in the case the average amplitude of fluctuations is large enough,
percolation occurs quickly and the two processes will have the same
rate, resulting in $f_{+}^{eq} = 0.5$.

The above argument is expected to apply only when $\Phi$ is localized
around the origin initially.  If it is localized around $\Phi=\Phi_{+}$
initially, on the contrary, we expect that the system will settle into
a state with $f_+^{eq}=1-I/(I+1)$ or $f_0^{eq}=I/(I+1)$, since the
potential we are using is symmetric.  In order to see what happens in
the case $\Phi$ is not localized around either minima, we have run
five simulations starting from a checkerboard made of $\Phi = 0$ and
$\Phi = \Phi_{+}$ using different realization of random numbers.  The
result is depicted in Fig.\ \ref{fig:half} for $\lambda=0.06,~\delta x
= 7.0$ and $N=64^3$, which shows that the system approaches to either
equilibrium state with $f_+^{eq}$ or $f_0^{eq}=I/(I+1)$, although it
takes much longer time to relax than in the cases with $\Phi=0$ or
$\Phi=\Phi_{+}$ initially.

This result implies that the configuration with $f_+=f_0=0.5$ is
unstable in this case even if it contains maximum number of
microscopic states.  Note that this configuration also costs more
energy of domain boundaries than any other configuration.  This is why
the system may relax to a configuration with $f_+ \neq f_0$.

Although these arguments are interesting in themselves, we must be
cautious with their interpretation, that is, it may not be directly
relevant to the actual dynamics of EWPT because our use of the
effective potential in the Langevin equation is not strictly
justifiable in the case field configuration becomes inhomogeneous.
The same warning also applies to Anderson's argument \cite{And}, who
claims that basic picture of phase transition through subcritical
bubbles is in contradiction with the second law of thermodynamics. 
This criticism is also based on an expression of the free energy which
is not strictly correct in inhomogeneous situations.

\section{Summary and Discussion} 

\label{sec:summary}

\indent

In the present paper we have performed a series of numerical
simulations of the Langevin equations toward understanding aspects of a
weakly first-order cosmological phase transition such as the
electroweak phase transition.

First we have confirmed that the simple Langevin equation
(\ref{eqn:simplelangevin}) with random Gaussian white noise
(\ref{eqn:whitenoise}) can reproduce thermal equilibrium state of a
massive non-selfinteracting scalar field, in particular, that the
correlation length is given by the inverse-mass scale independent of
the lattice spacing.  We have then applied the same technique to
one-loop improved effective potential of the Higgs field in the
electroweak theory.  Taking the coarse-graining scale or the lattice
spacing equal to the inverse-mass scale at $\phi = 0$, we have
confirmed that phase mixing does not occur at the critical temperature
for a small enough Higgs mass, consistent with the previous analytic
estimate of the amplitude of fluctuations \cite{DLHLL} \cite{Bet}
\cite{ERV}. At the same time, we have argued that in order to
reproduce the shape of the corresponding massive scalar correlation
function we should take $\delta x$ smaller. As a result of such
simulation we have found that the correlation function of the Higgs
field obtained numerically may damp at smaller scale depending on the
choice of $\delta x$, so it has proved that the so called
``correlation length'' or the inverse-mass scale at $\phi = 0$ is not
a good measure of a coarse-graining scale in the case the potential
contains a nontrivial interactions.  In this case we have also found
the final configuration of the simulation depends on the lattice
spacing severely, which is a manifestation of the fact that lattice
spacing serves as an ultraviolet cutoff of otherwise divergent theory.
Since the renormalization prescription of \cite{AG} does not work to
the temperature-dependent potential, we have tried to fix the lattice
spacing from a physical argument.

For this purpose we have reexamined derivation of the Langevin-like
equation in the literatures \cite{Mor} \cite{GR}. We stressed the
importance of the correlation of the noise terms which are the only
source of inhomogeneity in the system and so the simulation should be
done reflecting their properties.  Although the noise terms can be
derived from the perturbative non-equilibrium field theory, no
completely satisfactory derivation has been given of the other
ingredient of thermalization, namely, the dissipation terms. Hence we
made much of thermodynamics and determined them from the
fluctuation-dissipation relation.  Another difficulty of the equation
of motion from the perturbative effective action is that it contains
integral terms which are non-local in both space and time. Since it is
impossible to deal with them numerically, we have replaced them with
the derivative of the one-loop effective potential.  This procedure
may be justified if and only if the field configuration remains
homogeneous.  Since we set a homogeneous and static initial condition,
$\phi({\bf x},t)=\dot{\phi}({\bf x},t)=0,$ and what we are concerned
is if the system remains homogeneous, the above approximation is
sensible.  Note, however, that in the case phase mixing is manifest in
the final result, our simplified equation is no longer valid. Then the
one-loop approximation also breaks down at the same time and we would
have to deal with the full effective action, which is formidable now.

Keeping the above-mentioned limits of our approach in mind, let us
consider implication of the results of numerical calculations.  We
examined the effects of bosonic noises and fermionic noises separately.
In the former case the field remained practically homogeneous at least
for small enough values of $M_{H}$, and in the latter case phase
mixing was evident and our approximation broke down.  If only one
species of noises and the corresponding dissipation term are taken
into account, the strength of the noise and the dissipation do not
affect the final equilibrium configuration, as far as the fluctuation
dissipation relation is satisfied, although they do affect the
relaxation time scale. In the realistic case, both types of noises are
present and their amplitude is also important to determine
thermalization process of the system. Since the Yukawa coupling of the
top quark is larger than the square gauge coupling and fermionic noises
are more effective, we expect that the results of
subsection \ref{subsec:fermion} apply to the actual electroweak phase
transition.  In short, although gauge interaction plays the essential
role to induce the cubic term in the effective potential, the
non-equilibrium dynamics is dominated by fermionic interactions, and
as a result, the conventional picture of first-order phase transition
based on the one-loop potential is suspect.

So far we have traced the behavior of the expectation value of the
scalar field.  Although the equation of motion has been motivated from
quantum theory, as far as we concentrate on an expectation value, we
must make sure that the effect of quantum uncertainty is sufficiently
small. This condition is satisfied if the number of quanta contained
in one lattice volume is much larger than unity \cite{AHH} \cite{And}.
The simulations with a lattice spacing equal to the fundamental
correlation length of the fermionic noise, $\beta/(2\pi)$, does not
satisfy this constraint. Our conclusion, however, remain intact since
at the critical value of $\delta x=6.0$ in Fig.\ \ref{fig:lambda}, one
lattice volume already contains more than 40 quanta, much larger than
unity.

In the light of the above results and from the fact that all the
previous literatures claiming that usual nucleation picture in the
homogeneous background applies to the electroweak phase transition
have estimated fluctuation on too large a spatial scale, it is now
evident that the conventional picture with one-loop effective
potential does not work. On the other hand, in order to clarify real
dynamics of the phase transition we must say that much work is to be
done including derivation of correct equation of motion of the order
parameter which should replace the crude phenomenological equation
employed here.

\subsection*{Acknowledgments}

We would like to thank S.\ Mukohyama for discussion.  MY is grateful
to Professor K. Sato for his continuous encouragement and to M.\
Morikawa, T.\ Shiromizu, and, M.\ Kawasaki for their useful comments.
This work was partially supported by the Japanese Grant in Aid for
Scientific Research Fund of the Ministry of Education, Science, Sports
and Culture Nos.\ 07304033(JY), 08740202(JY), and 09740334(JY).



\begin{figure}[htb]
\makeatletter
\def\fnum@figure{\figurename~\thefigure (a)}
\makeatother
  \begin{center} 
\leavevmode\psfig{figure=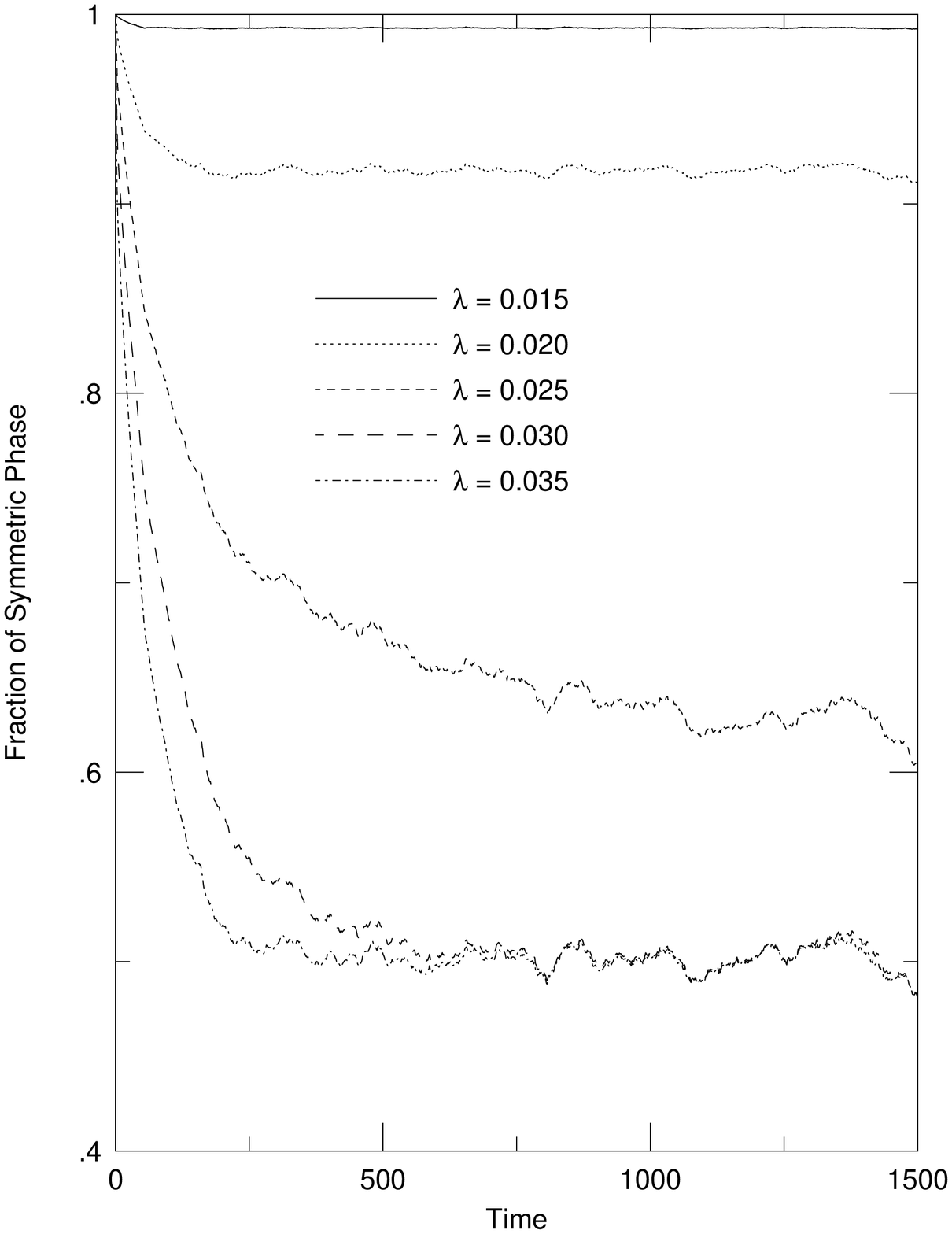,height=20cm}
  \end{center}
     \caption{Fraction of the symmetric phase, $f_{0}$, for $\delta
     x=1.0$.}
\label{fig:normal}
\end{figure}

\addtocounter{figure}{-1}
\begin{figure}[htb]
\makeatletter
\def\fnum@figure{\figurename~\thefigure (b)}
\makeatother
  \begin{center} 
\leavevmode\psfig{figure=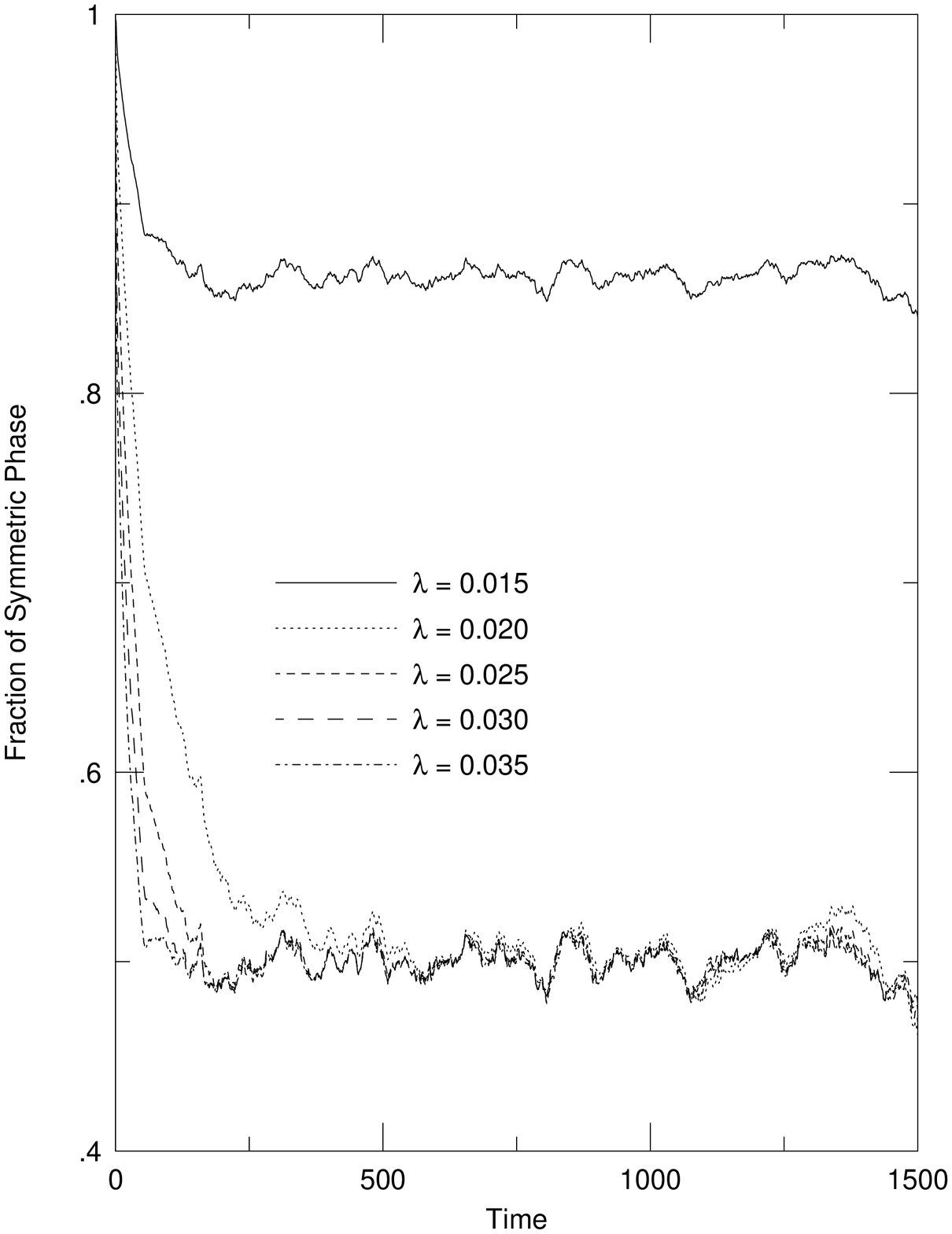,height=20cm}
  \end{center}
     \caption{Fraction of the symmetric phase, $f_{0}$, for $\delta
     x=0.5$.}
\label{fig:small}
\end{figure}

\begin{figure}[htb]
\makeatletter
\def\fnum@figure{\figurename~\thefigure (a)}
\makeatother
  \begin{center} 
\leavevmode\psfig{figure=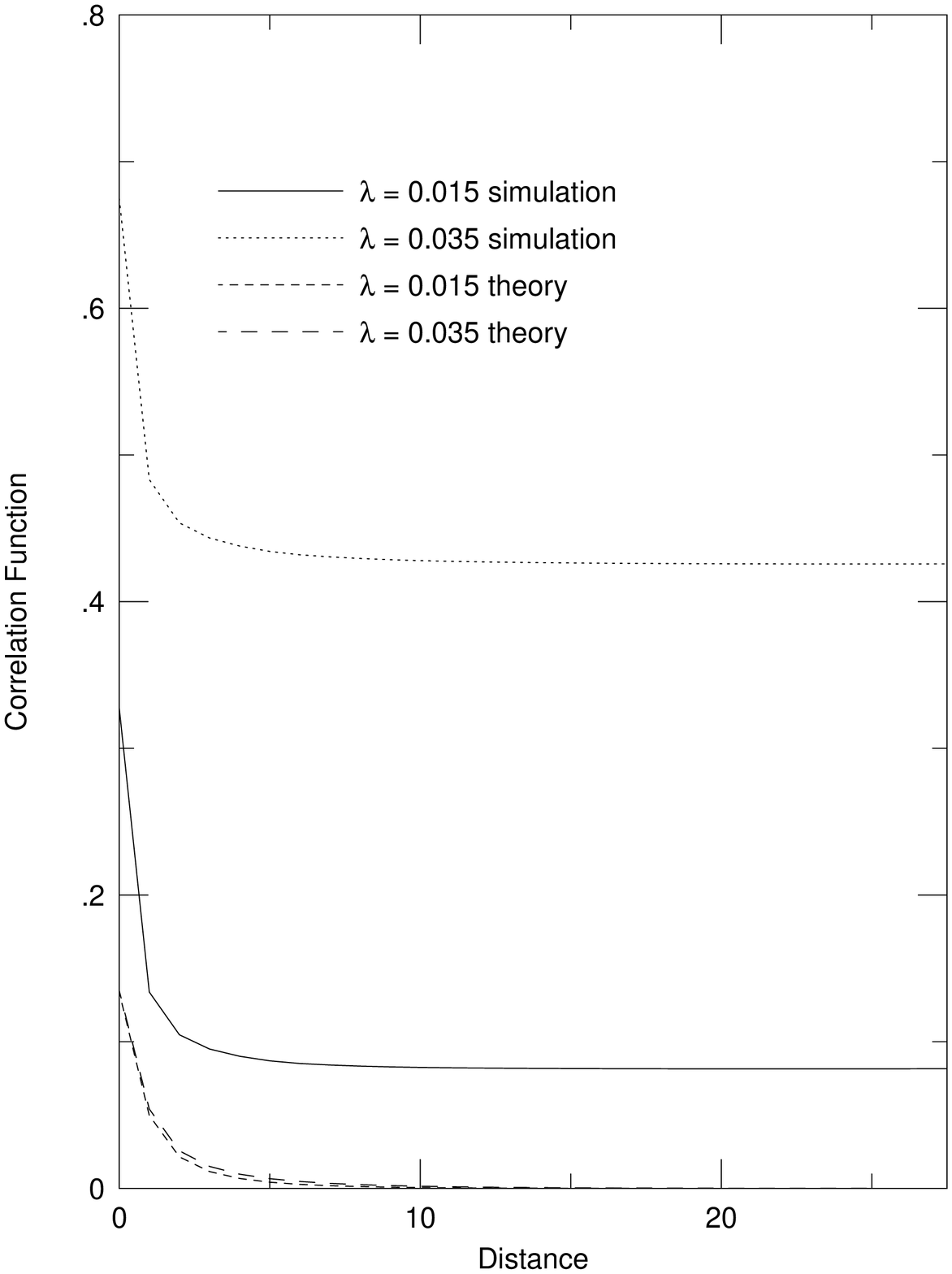,height=20cm}
  \end{center}
     \caption{Correlation function $\la \Phi(\vect x)\Phi(\vect y)
     \ra$ for the scalar field at with the one-loop electroweak
     potential ($\delta x=1.0$).}
\label{fig:cnormal}
\end{figure}

\addtocounter{figure}{-1}
\begin{figure}[htb]
\makeatletter
\def\fnum@figure{\figurename~\thefigure (b)}
\makeatother
  \begin{center} 
\leavevmode\psfig{figure=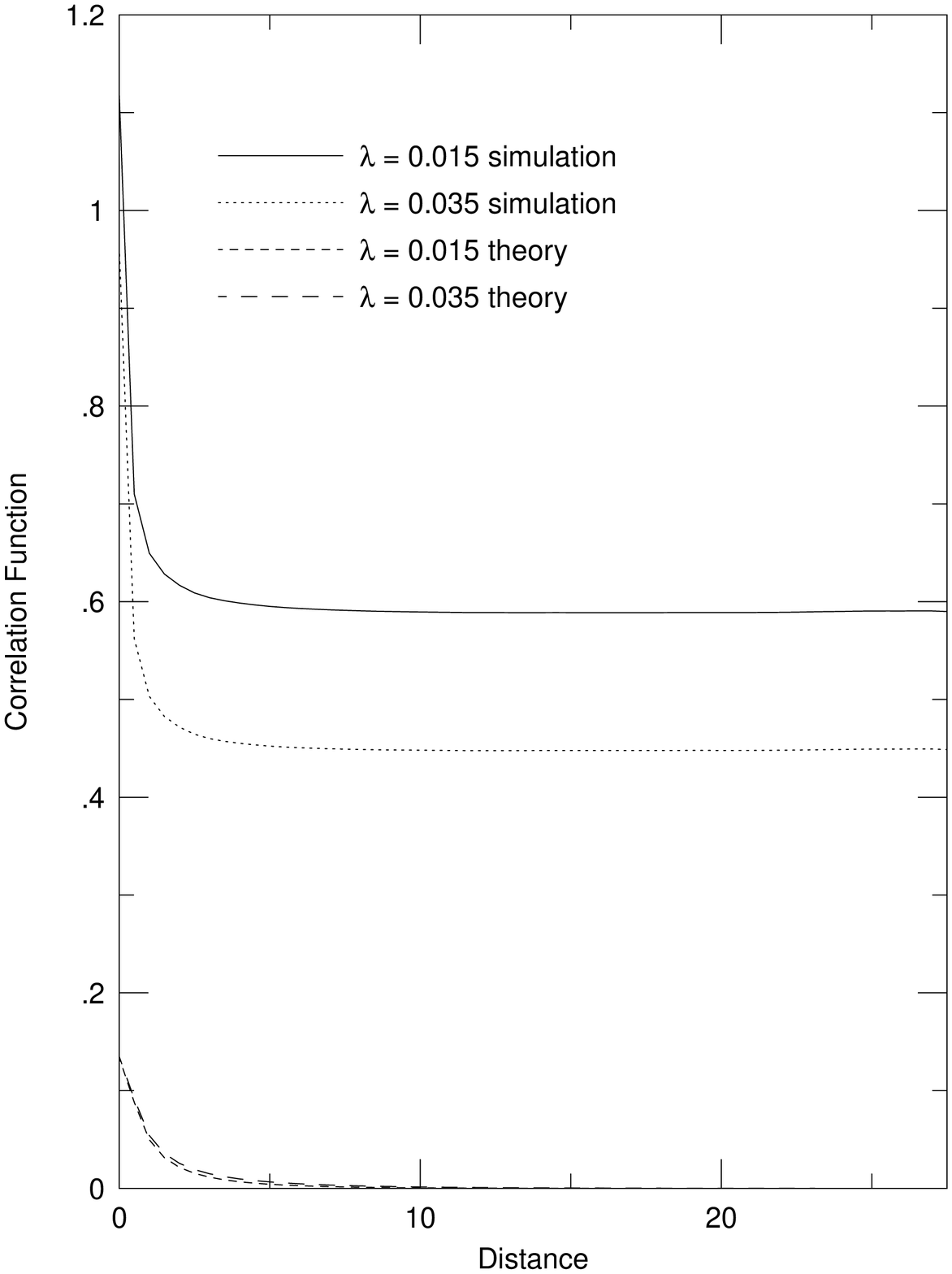,height=20cm}
  \end{center}
     \caption{Correlation function $\la \Phi(\vect x)\Phi(\vect y)
     \ra$ for the scalar field at with the one-loop electroweak
     potential ($\delta x=0.5$).}
\label{fig:csmall}
\end{figure}

\begin{figure}[htb]
\begin{center}
\leavevmode\psfig{figure=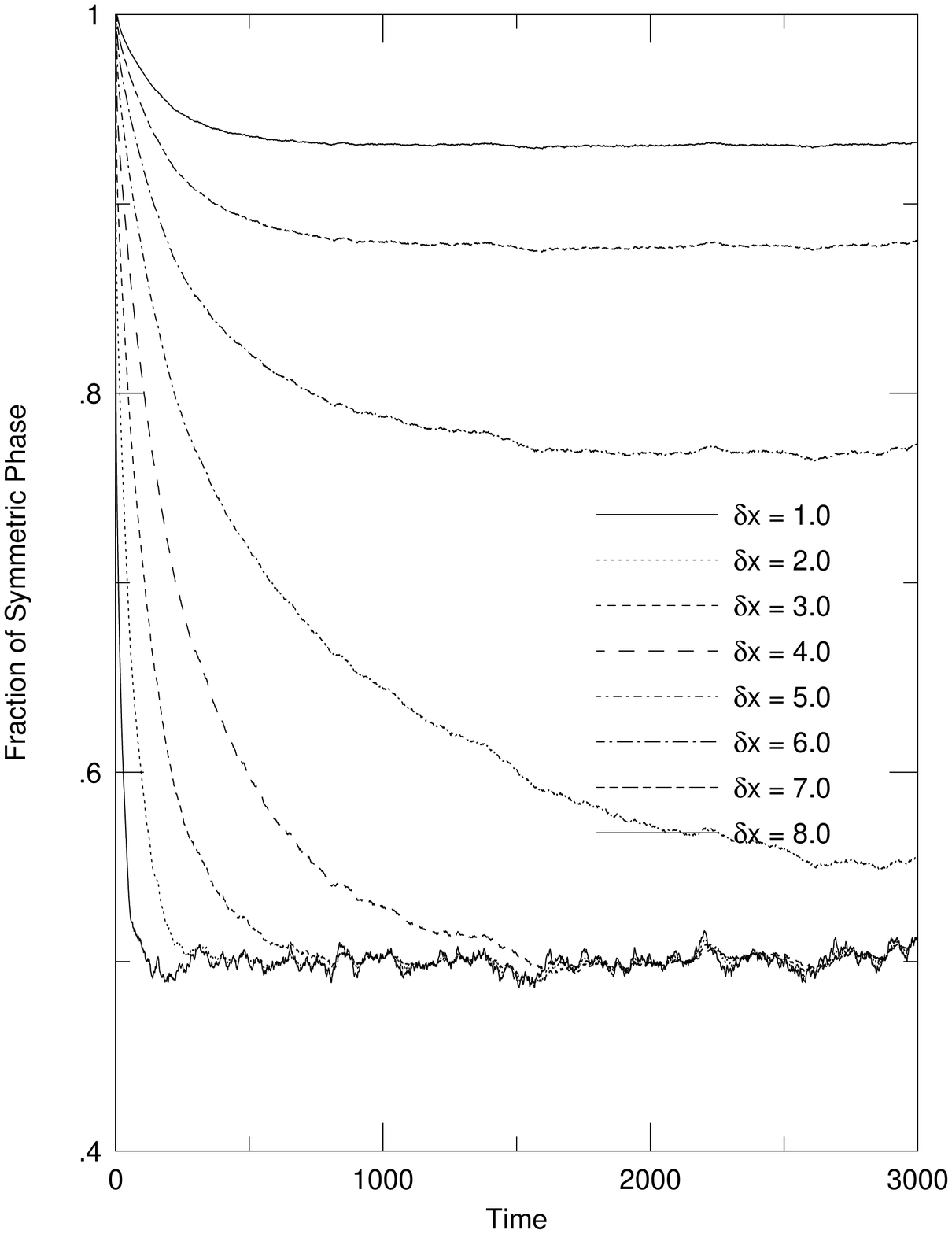,height=20cm}
\end{center}
\caption{Fraction of the symmetric phase, $f_{0}$, for $\delta x$ = 1.0, 
2.0, 3.0, 4.0, 5.0, 6.0, 7.0, and 8.0 from the bottom to the top with
$\lambda=0.060$.} 
\label{fig:lambda}
\end{figure}

\begin{figure}[htb]
\begin{center}
\leavevmode\psfig{figure=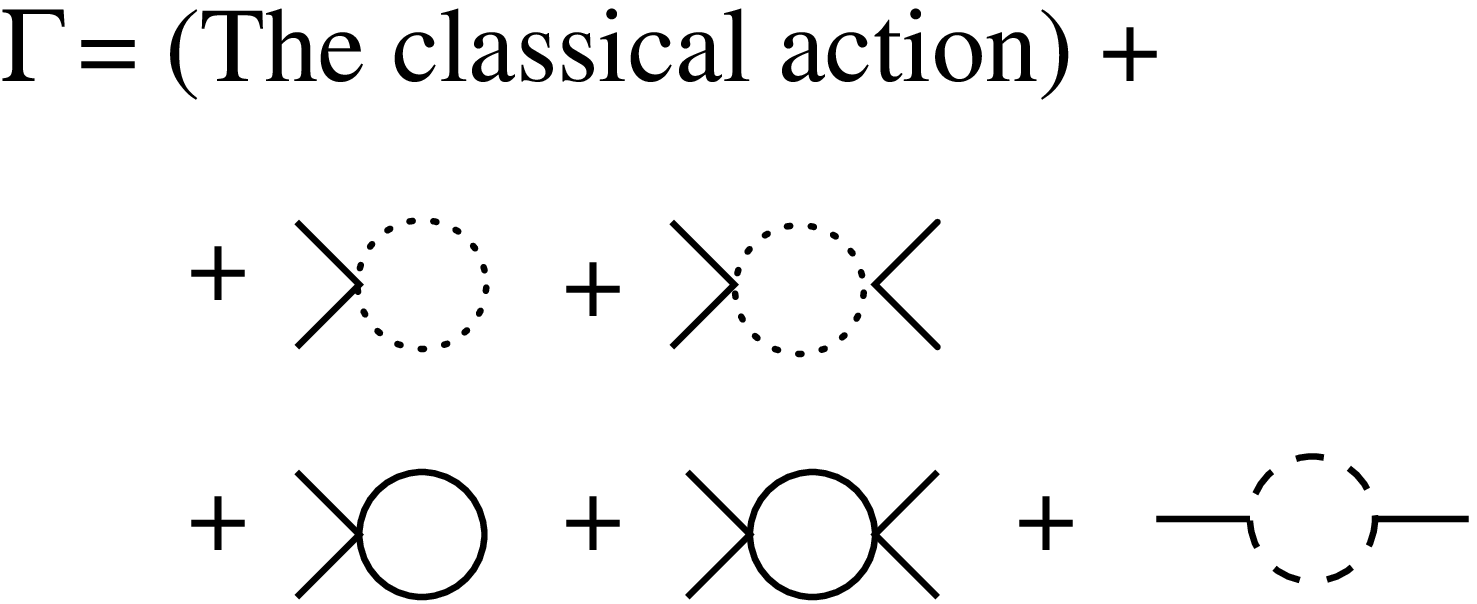,width=17cm}
\end{center}
\caption{Contribution to $\Gamma$ up to one loop order and
$\CO(\lambda^2, g^4,f^2)$. Solid lines represent $\phi$ fields, dotted 
lines $\chi$ fields, and dashed lines $\psi$ fields.}  
\label{fig:one}
\end{figure}

\begin{figure}[htb]
\begin{center}
\leavevmode\psfig{figure=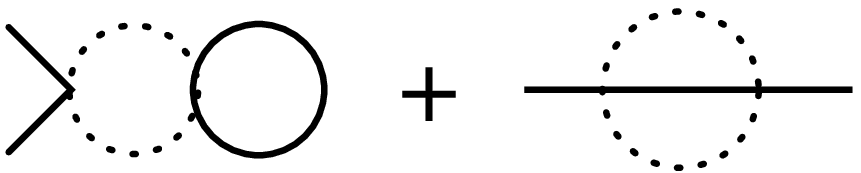,width=17cm}
\end{center}
\caption{Contribution to $\Gamma$ with two loops order and
$\CO(\lambda^2, g^4,f^2)$.} 
\label{fig:two}
\end{figure}

\begin{figure}[htb]
\begin{center}
\leavevmode\psfig{figure=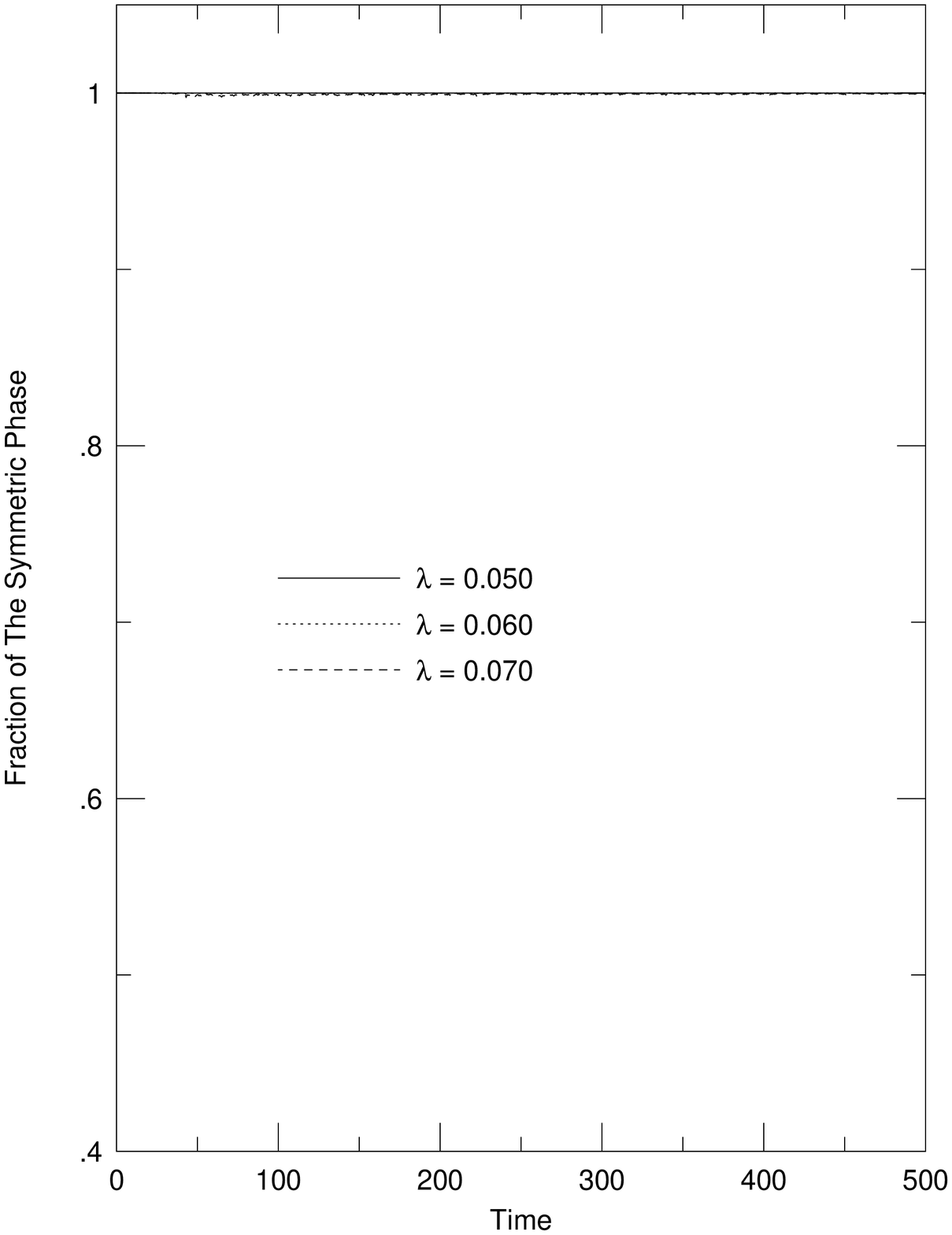,height=20cm}
\end{center}
\caption{Fraction of the symmetric phase, $f_{0}$, for the massless
bosonic noise.} 
\label{fig:boson}
\end{figure}

\begin{figure}[htb]
\begin{center}
\leavevmode\psfig{figure=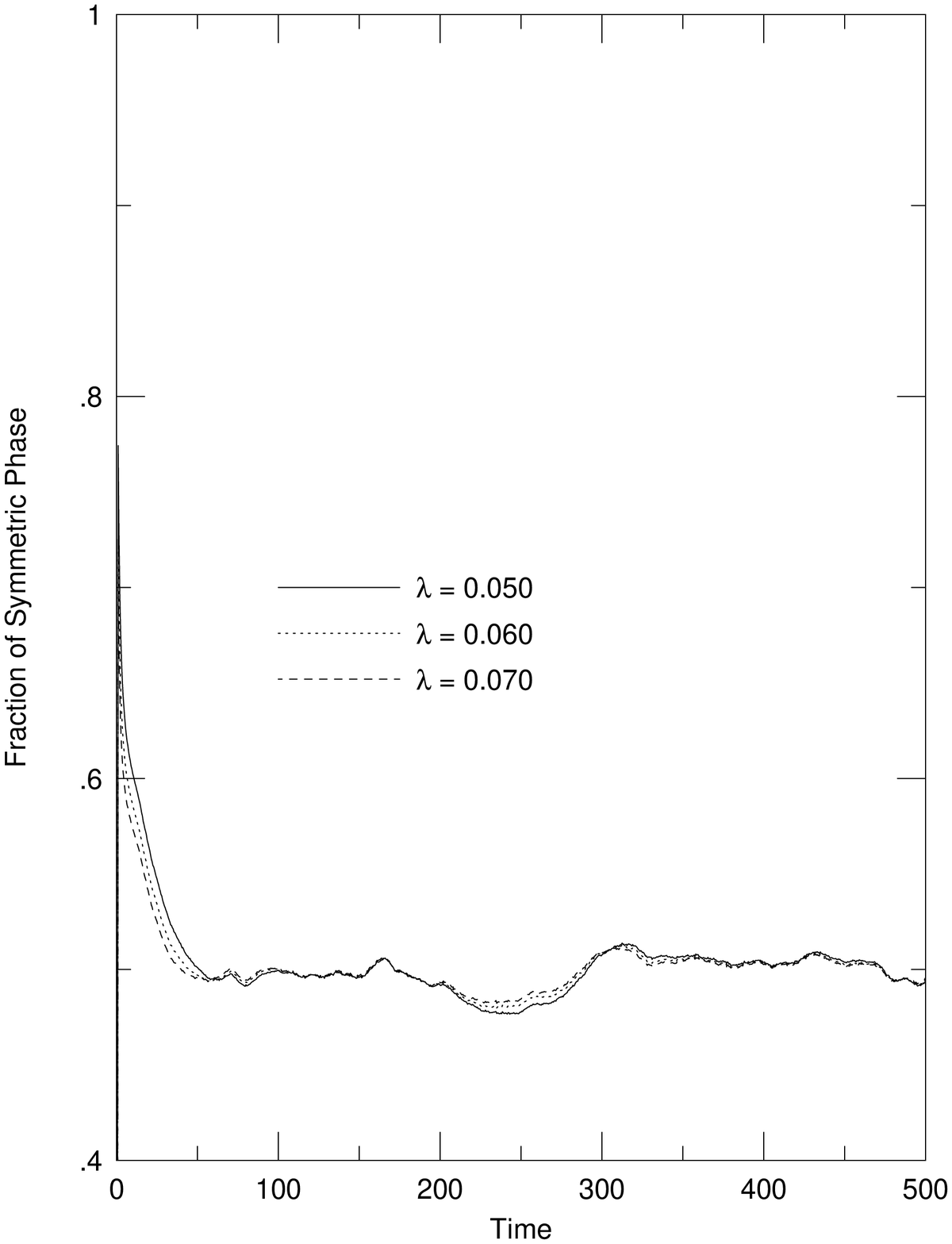,height=20cm}
\end{center}
\caption{Fraction of the symmetric, $f_{0}$, phase for the massless
fermionic noise.}
\label{fig:fermion}
\end{figure}

\begin{figure}[htb]
\makeatletter
\def\fnum@figure{\figurename~\thefigure (a)}
\makeatother
  \begin{center}
\leavevmode\psfig{figure=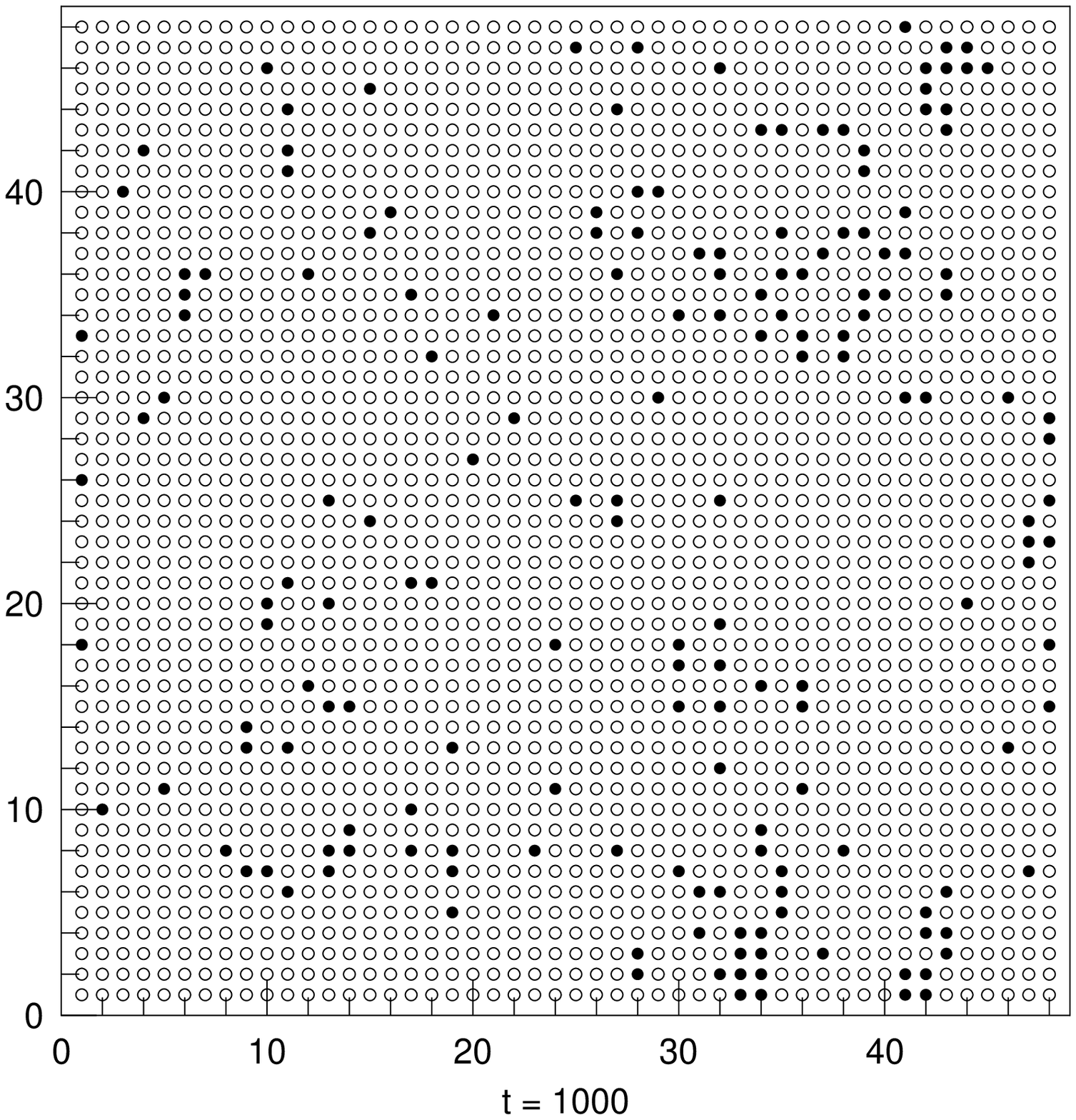,height=20cm}
  \end{center}
     \caption{Snap shot in a plain at $t = 1000$ of the simulation
     with $\delta x = 1.0$ and $\lambda = 0.02$. Black dots represent
     asymmetric-state bubbles, while white ones a symmetric-state
     background.}
\label{fig:t1000}
\end{figure}

\addtocounter{figure}{-1}
\begin{figure}[htb]
\makeatletter
\def\fnum@figure{\figurename~\thefigure (b)}
\makeatother
  \begin{center}
\leavevmode\psfig{figure=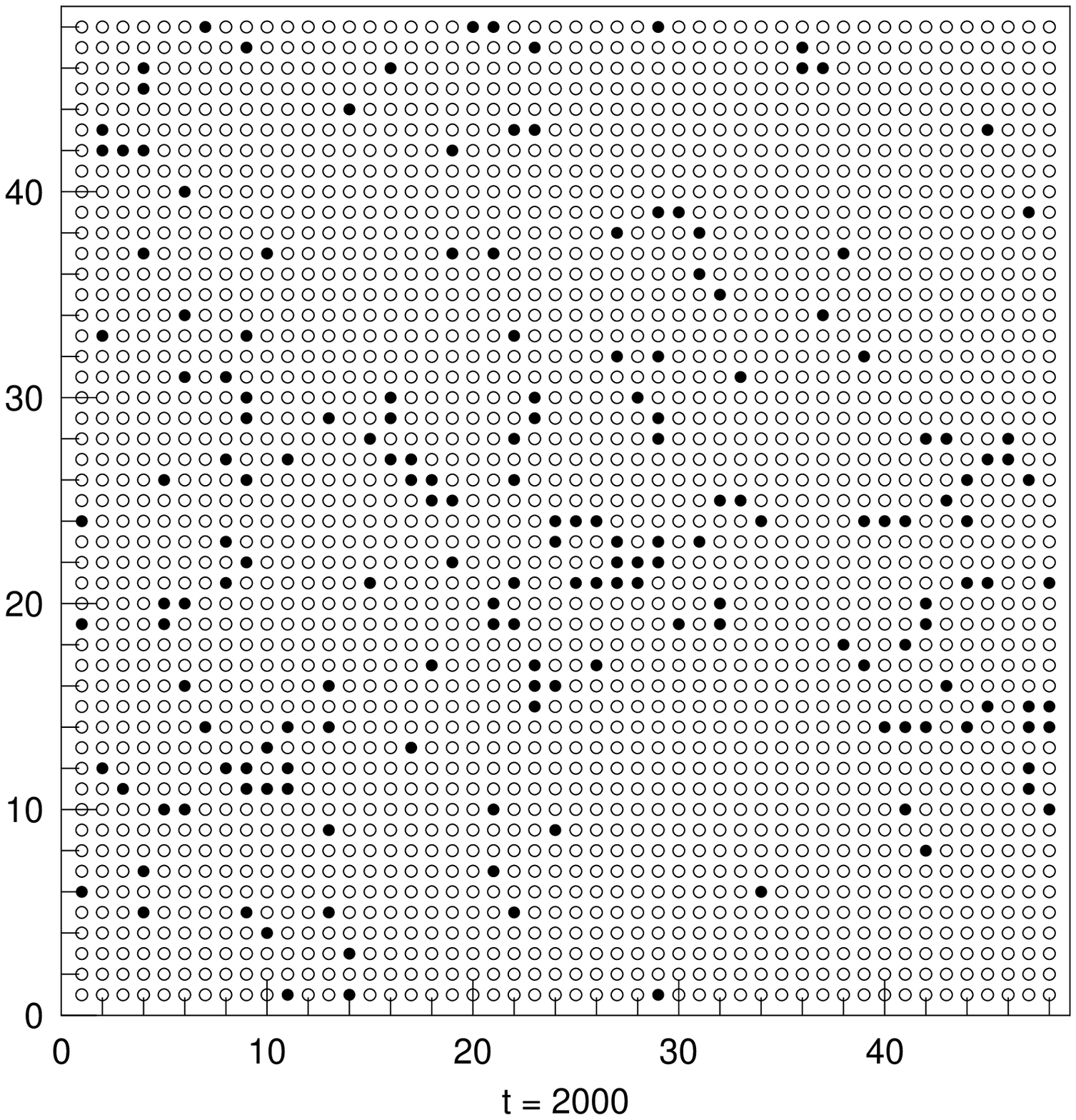,height=20cm}
  \end{center}
     \caption{Snap shot at $t = 2000$.}
\label{fig:t2000}
\end{figure}

\addtocounter{figure}{-1}
\begin{figure}[htb]
\makeatletter
\def\fnum@figure{\figurename~\thefigure (c)}
\makeatother
  \begin{center}
\leavevmode\psfig{figure=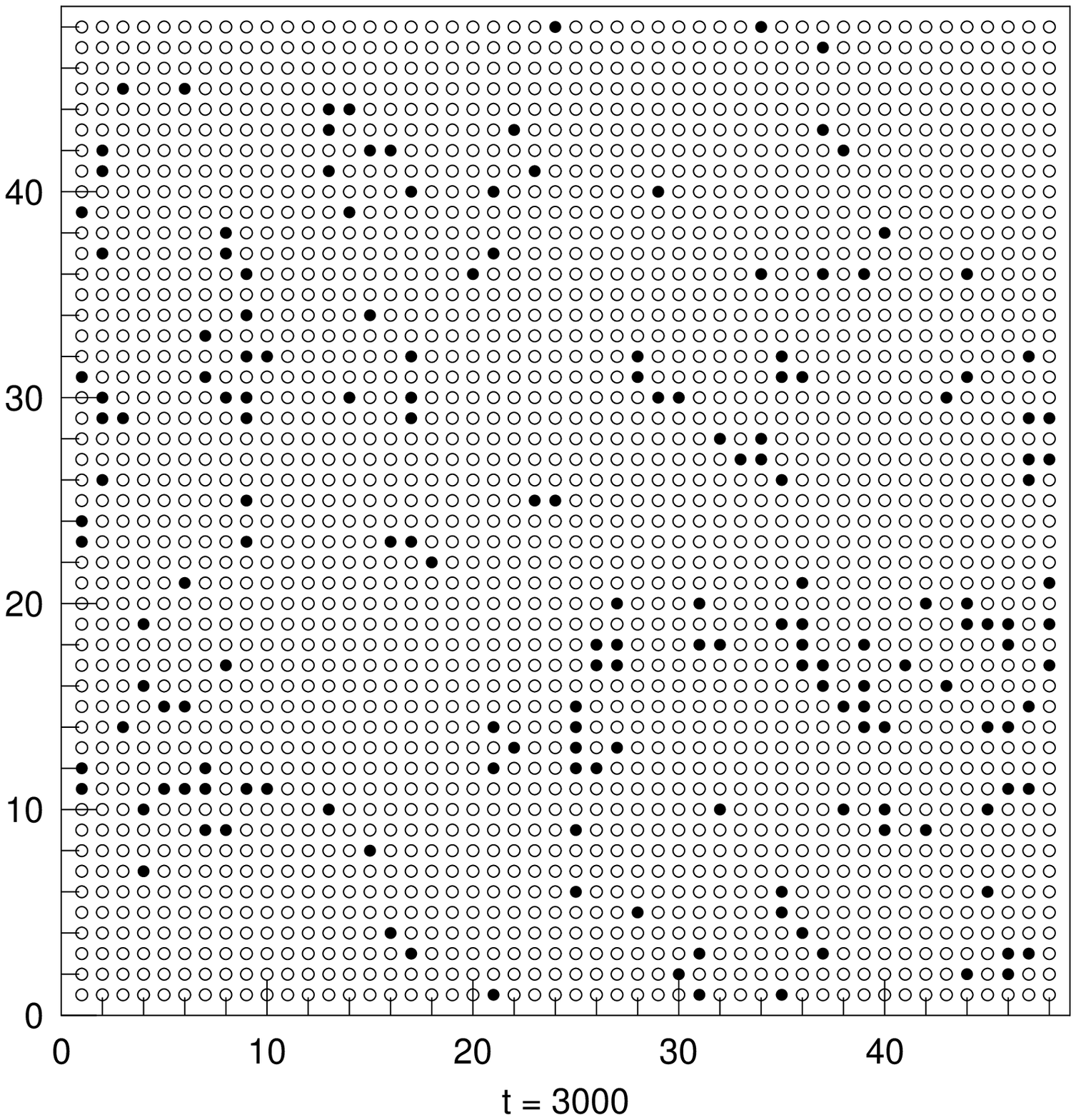,height=20cm}
  \end{center}
     \caption{Snap shot at $t = 3000$.}
\label{fig:t3000}
\end{figure}

\addtocounter{figure}{-1}
\begin{figure}[htb]
\makeatletter
\def\fnum@figure{\figurename~\thefigure (d)}
\makeatother
  \begin{center}
\leavevmode\psfig{figure=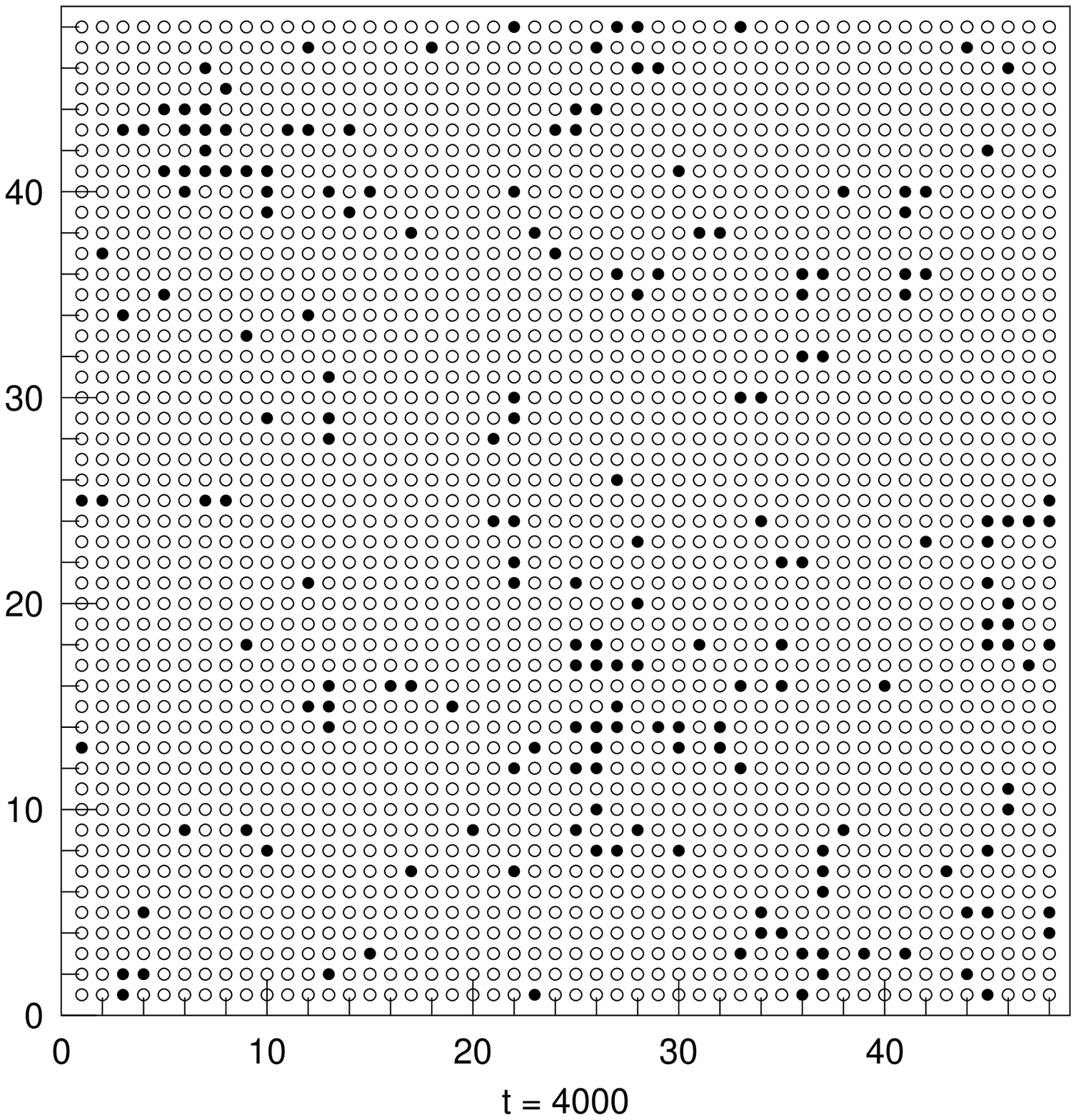,height=20cm}
  \end{center}
     \caption{Snap shot at $t = 4000$.}
\label{fig:t4000}
\end{figure}

\addtocounter{figure}{-1}
\begin{figure}[htb]
\makeatletter
\def\fnum@figure{\figurename~\thefigure (e)}
\makeatother
  \begin{center}
\leavevmode\psfig{figure=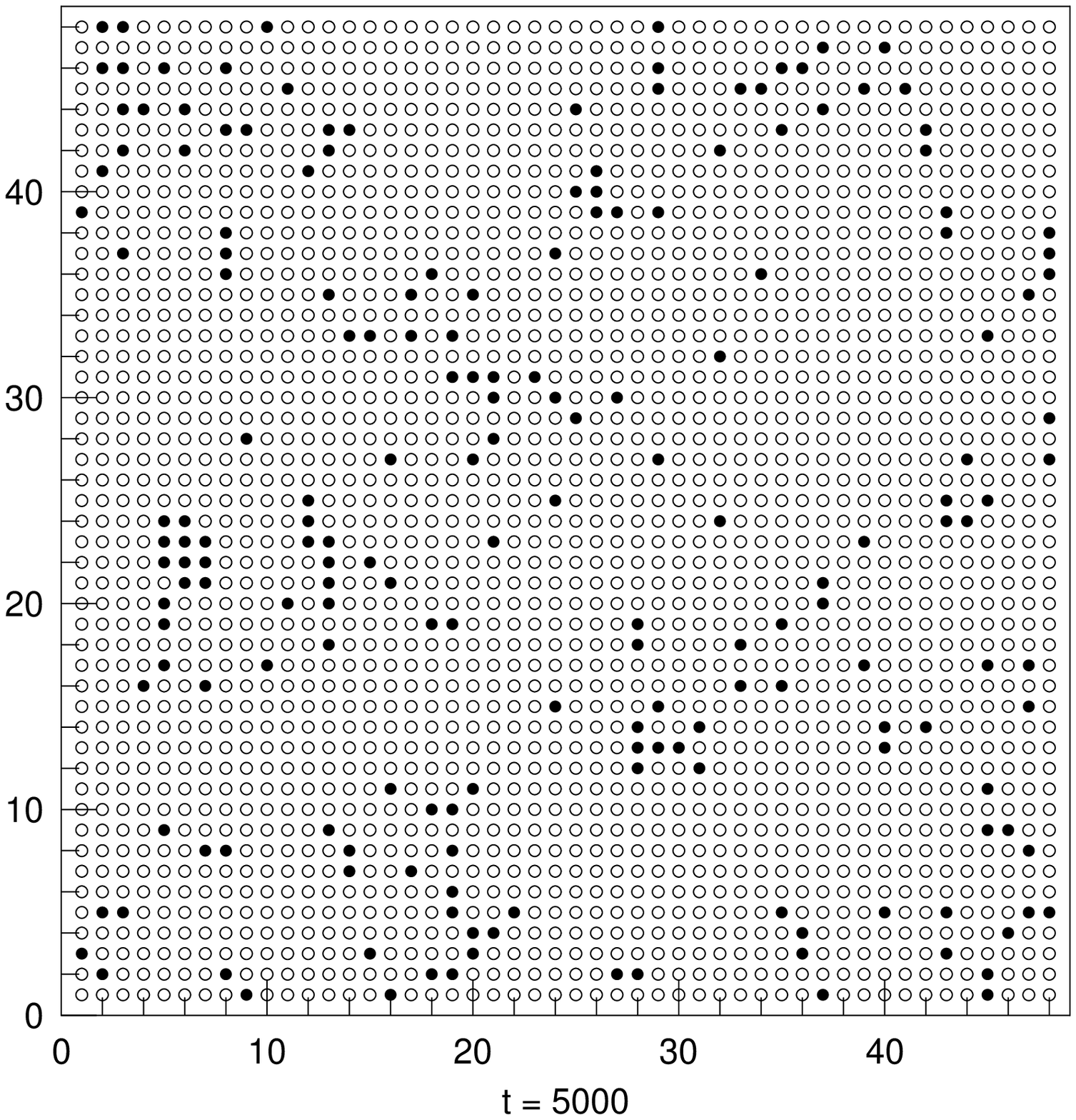,height=20cm}
  \end{center}
     \caption{Snap shot at $t = 5000$.}
\label{fig:t5000}
\end{figure}

\begin{figure}[htb]
\begin{center}
\leavevmode\psfig{figure=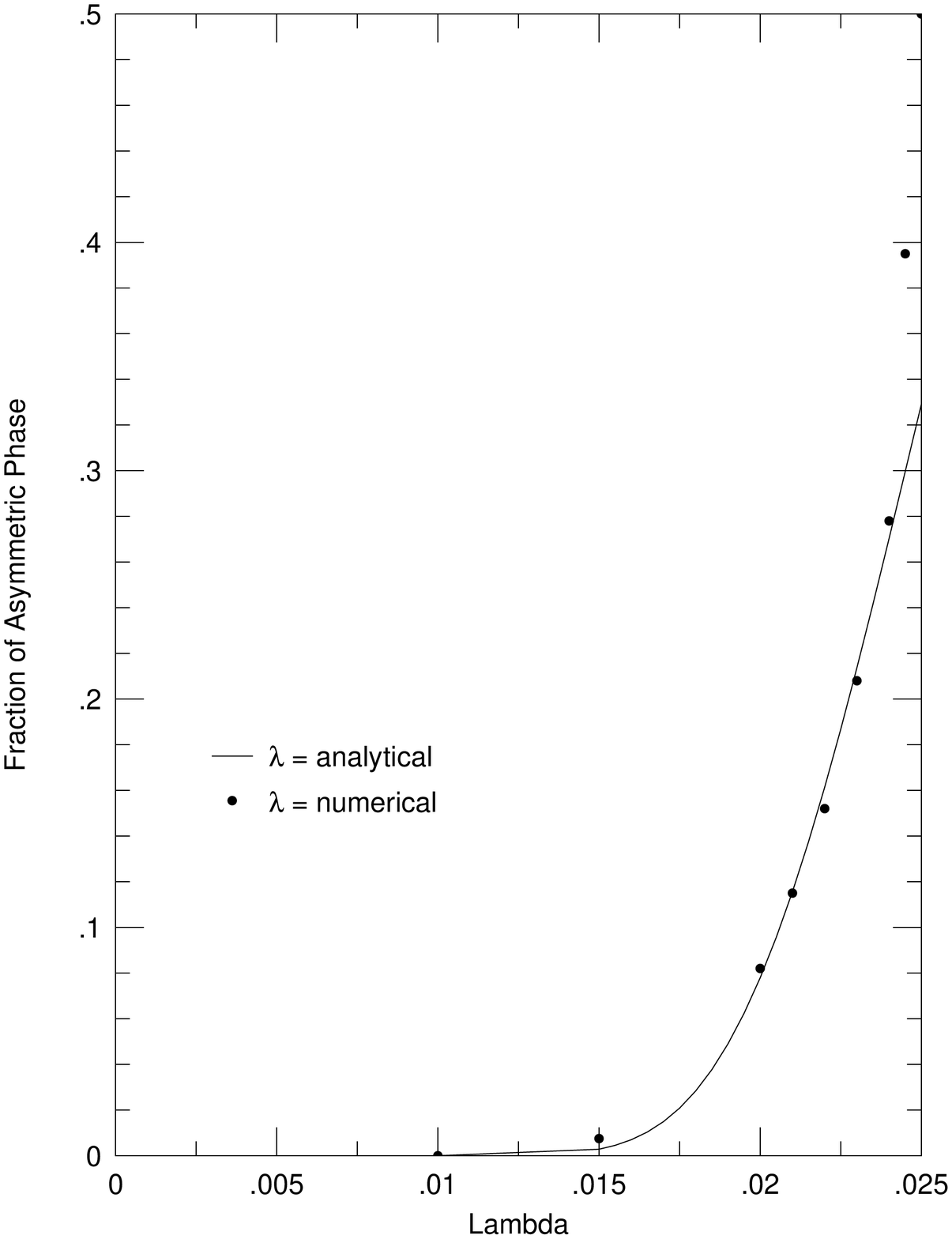,height=20cm}
\end{center}
\caption{Comparison of analytical and numerical results.}
\label{fig:analytical}
\end{figure}

\begin{figure}[htb]
\begin{center}
\leavevmode\psfig{figure=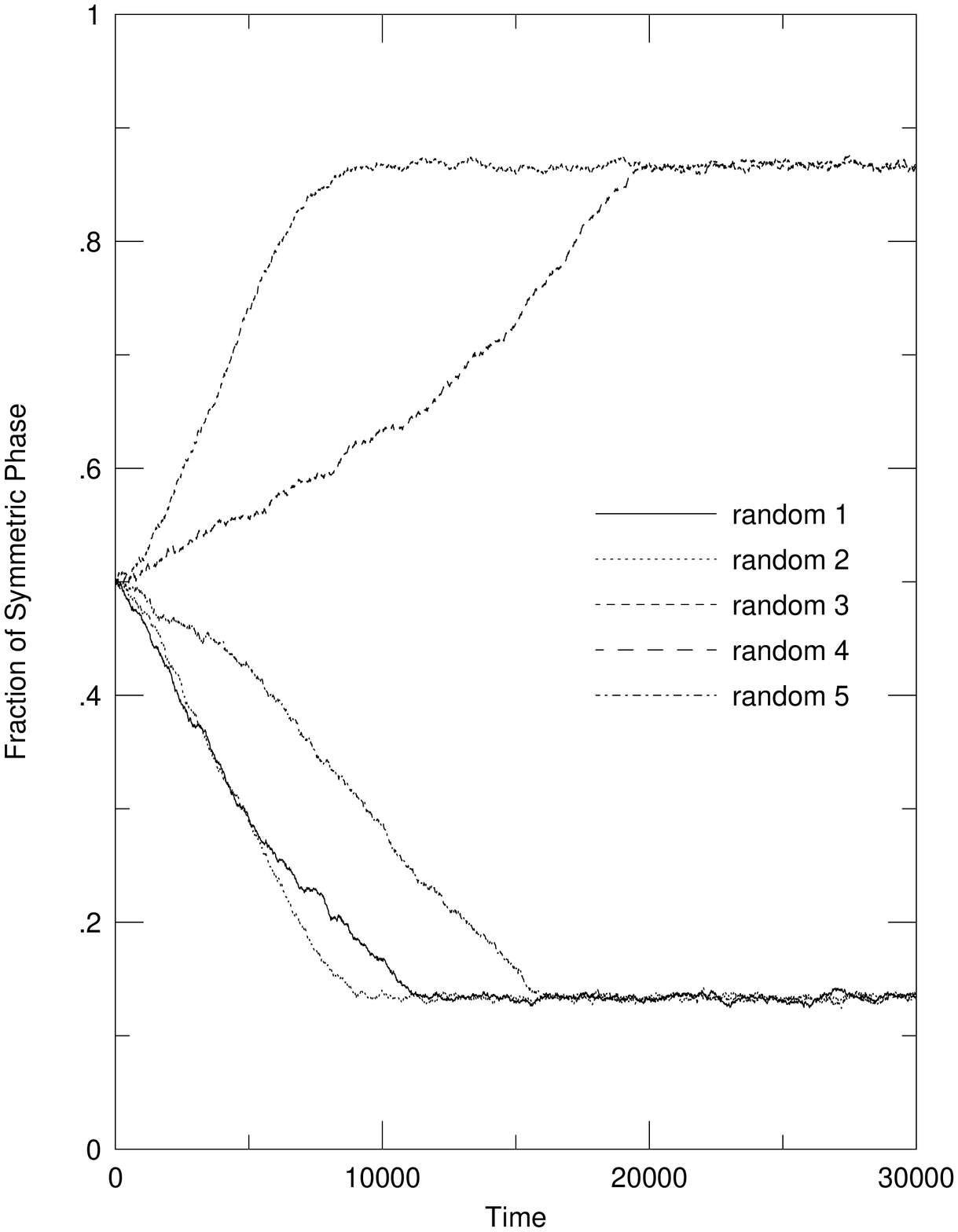,height=20cm}
\end{center}
\caption{Fraction of the symmetric, $f_{0}$ starting from a checkerboard.}
\label{fig:half}
\end{figure}

\end{document}